\RequirePackage{fix-cm}
\documentclass[twocolumn,epjc3]{svjour3}       
\smartqed  
\usepackage{graphicx}

\usepackage{cite}
\usepackage[T1]{fontenc} 
\usepackage{color}

\usepackage{subfig}
\usepackage{amsmath}
\usepackage{textcomp}
\usepackage{hyperref}

\usepackage{float}

\newcommand{\ba}{\begin{array}}
\newcommand{\ea}{\end{array}}
\newcommand{\bd}{\begin{displaymath}}
\newcommand{\ed}{\end{displaymath}}
\newcommand{\bsube}{\begin{subequation}}
\newcommand{\esube}{\end{subequation}}
\newcommand{\bea}{\begin{eqnarray}}
\newcommand{\eea}{\end{eqnarray}}
\newcommand{\bal}{\begin{align}}
\newcommand{\ealign}{\end{align}}
\newcommand{\eal}{\end{align}}
\newcommand{\ben}{\begin{enumerate}}
\newcommand{\een}{\end{enumerate}}
\newcommand{\nn}{\nonumber}
\newcommand{\dis}{\displaystyle}

\newcommand{\gev}{\; {\rm GeV} }
\newcommand{\tev}{\; {\rm TeV} }

\newcommand{\lsim}{\:\raisebox{-0.5ex}{$\stackrel{\textstyle<}{\sim}$}\:}

 \usepackage[normalem]{ulem}
 
 \definecolor{darkgreen}{cmyk}{1,0,1,0.4}
 \definecolor{pink}{cmyk}{0.4,1,0.3,0}

\begin{document} 
\title{Signals of Leptophilic Dark Matter at the ILC}

\author{Sukanta Dutta\thanksref{addr1,e1} \and Bharti Rawat \thanksref{addr2,e2} \and Divya Sachdeva \thanksref{addr2,e3}}
\thankstext{e1}{e-mail: Sukanta.Dutta@gmail.com}
\thankstext{e2}{Corresponding~Author, \hfill\break \hspace*{3pt} e-mail: bhartirawat87@gmail.com}
\thankstext{e3}{e-mail: divyasachdeva951@gmail.com}
\institute{
\emph{SGTB Khalsa College, University of
Delhi. Delhi-110007. India.}\label{addr1}
\and
\emph{Department of Physics and Astrophysics, University of
Delhi. Delhi-110007. India.}\label{addr2}
}


\date{}

\maketitle

\abstract{ Adopting a model independent approach, we constrain the various effective 
 interactions of leptophilic DM particle with the visible world from the WMAP and Planck data. The thermally averaged indirect DM annihilation cross-section and the DM-electron direct-detection cross-section for such a DM candidate are observed to be consistent with the respective experimental data.  
\par We study the  production of cosmologically allowed leptophilic DM in association with $Z\, (Z\to f\bar f)$, $f\equiv q,\,e^-,\, \mu^-$ at the ILC. We perform the $\chi^2$ analysis and compute the 99\% C.L. acceptance contours in the $m_\chi$ and $\Lambda$ plane from the two dimensional differential distributions of various kinematic observables obtained after employing parton showering and hadronization to the simulated data. We observe that the dominant hadronic channel provides the best kinematic reach of 2.62 TeV ($m_\chi$ = 25 GeV), which further improves to 3.13 TeV for polarized beams at $\sqrt{s} = 1$ TeV and an integrated luminosity of 1 ab$^{-1}$. }

 \section{Introduction}
\label{sec:intro}

Dark Matter provides the most compelling explanation for many
cosmological and astrophysical observations which defy an understanding in
terms of luminous matter alone. However, in the absence of any
direct evidence, the existence of such matter has, rightly, been
questioned and several alternatives proposed, including the
modification of gravity at large distance
scales. It should be appreciated, though, that starting with orbital
velocities of stars in a galaxy or those of the galaxies themselves in
a cluster~\cite{galaxy:RubinFord}, gravitational
lensing~\cite{grav_lensing,weaklensing}, the dynamics of galactic
(cluster) collisions~\cite{Clowe:2006eq} etc., the observations span a
wide range of distance scales, and no single simple modification of
gravitation can explain all, whereas dark matter (DM) can and does
play an important role in understanding the data. Similarly, the
fitting of the cosmological observables
~\cite{Hinshaw:2012aka,Abadi:2002tt,Ade:2015xua}, requires that DM
contributes about 23\% to the energy budget of the universe, in
contrast to only 4\% contained in baryonic matter. Finally,
post-inflation, perturbations in the DM distribution, along with the
gravitational perturbations, are supposed to have provided the seed
for large-scale structure formation in the
universe~\cite{Abadi:2002tt}. The constraints from the latter are
rather strong. Indeed, while neutrinos would have been logical
candidates for DM within the Standard Model (SM), a large energy
component in neutrinos would have disrupted structure formation. For
example, a recent study claims that if the equation of state for the
DM be parametrized as $p = w \, \rho$, then the combined fitting of
the cosmic microwave background, the baryon acoustic oscillation data
and the Hubble telescope data restricts $-9.0 \times 10^{-3} < w < 2.4
\times 10^{-3}$, thereby clearly preferring a cold and dusty
DM~\cite{Thomas:2016iav}.

With the DM particle, by definition, not being allowed to have either
electromagnetic\footnote{While DM with ultra-suppressed electromagnetic
couplings has been considered~\cite{Overduin:2004sz}, such models 
are extremely ungainly and are unlikely to survive closer 
scrutiny.} or strong interaction, the most popular\footnote{An alternative
could be an ultralight boson, such as that in Ref.\cite{Hui:2016ltb}. However, such a solution would, essentially, be untestable in the laboratory 
in the foreseeable future. Similar is the case for an ultralight gravitino DM candidate~\cite{Moultaka:2006su,Dutta:2015ega}} candidate is
the weakly interacting massive particle (WIMP). Recently, the authors of reference~\cite{Davoudiasl:2017jke} have proposed the cosmologically allowed super-light fuzzy dark matter of the order of 10$^{-22}$ eV which can possibly be  testable in terrestrial neutrino oscillation experiments. And while its exact
nature is unknown, several theoretical scenarios such as multi-Higgs
models, supersymmetry, extra-dimensional theories, little Higgs
models, left-right symmetric models all naturally admit viable candidates.
Consequently, one of the most challenging tasks today is to identify
the nature of the DM particle~\cite{Feng:2010gw}. This, in principle, could be done in
three kind of experiments. {\em Direct detection} can be achieved by
setting up apparatii (very often ultra-cold bolometric devices) that
would register the scattering of a DM particle off the detector
material.  While the DAMA experiment~\cite{Bernabei:2013xsa} did indeed claim the existence of
a DM particle of mass $\sim 60\gev$ from observed seasonal variation
in the detector signal (originating, possibly, from a varying DM wind as
the Earth traverses its path), subsequent experiments like CoGeNT ~\cite{Aalseth:2012if}, CRESST-II ~\cite{Angloher:2016rji}, XENON100 ~\cite{Aprile:2016wwo},  PandaX-II ~\cite{Yang:2016odq,Tan:2016zwf} and LUX ~\cite{Akerib:2016vxi} have not
validated this; rather, they have only served to impose bounds in the plane
described by the DM particle mass and its coupling to nucleons. {\em Indirect detection} experiments, largely satellite-based, depend on
the annihilation of a pair of DM particles in to SM particles which
are, subsequently, detected. Although there, occasionally, have been
claims of anomalies in the data, unfortunately the experiments have
failed to validate each other's positive sightings, resulting, once
again, in further constraints\cite{Fermi-LAT:2016uux,Adriani:2008zr,Aguilar:2013qda}. 
Finally, we have the {\em collider experiments}, wherein excesses (over the SM expectations) in final
states with large missing momentum are looked for. It should be
realized, though, that even if such an excess is established, a DM 
explanation would still only be an hypothesis, for
the only statement that can be made with certainty is that the
produced neutral and colour-singlet particle is stable over the
detector dimensions.

It is the last mentioned approach that we assume in this paper.  While
an investigation of the nature of the DM particles needs an
understanding of the underlying physics, we adopt, instead, a model
independent approach. Eschewing the details of the underlying dynamics,
we begin by postulating a fermionic DM particle, and consider
four-fermi operators involving these and the SM fermions. The relative
strengths of these operators would, of course, be determined by the
underlying dynamics. We assume that the operators involving quark currents are
subdominant, as could happen, for example, if the 
dynamics, at a more fundamental level, involved a leptophilic boson. This immediately negates the constraints
from the direct detection experiments~\cite{Aprile:2016wwo,Yang:2016odq,Tan:2016zwf,Akerib:2016vxi} (as the dominant interactions
therein are with the nucleons) as also bounds from the LHC~\cite{Bhattacherjee:2012ch,Goodman:2010ku,Carpenter:2012rg,Petrov:2013nia,Carpenter:2013xra,Berlin:2014cfa,Lin:2013sca,Bai:2010hh,Birkedal:2004xn,Gershtein:2008bf,Fox:2011fx,Bai:2012xg,Crivellin:2015wva,Petriello:2008pu,Feng:2005gj}.

Such an assumption also alters the conclusions (on the
interrelationship between the DM-mass and the coupling strength) drawn
from the deduced relic density.  Starting with the Boltzmann equations
describing the evolution of the particle densities, we derive the
constraints on the same. As for the search strategies at a linear
collider, the attention of the community, so far, has been largely commandeered
by the final state comprising a single photon accompanied by missing
energy~\cite{Chae:2012bq,Dreiner:2012xm,Yu:2013aca,Bartels:2007cv}. While this continues to be an important channel, we augment
this by other channels that are nearly as sensitive. Furthermore, we 
consider some novel effects of beam polarization.

 \section{Fermionic Dark Matter : A mini review}
\label{sec:fermionicdm}

As we have already mentioned, rather than considering the intricacies 
of particular models, we adopt the more conservative yet powerful 
concept of an effective field theory. Assuming that we have a 
single-component DM in the shape of a Dirac fermion $\chi$, 
we consider the least-suppressed (namely, dimension-six)
operators involving a $\chi$-current and a SM fermion-current
\footnote{We do not consider operators involving the SM bosons
as they play only a subservient role in both direct and indirect 
detection.}. This simplifying assumption rules out the possibility 
of resonances and co-annihilations affecting, significantly, the
relic abundance of dark matter particle \cite{Griest:1990kh}. 
Furthermore, to reduce the number of possible operators, as also 
not induce flavour changing neutral current processes, we restrict 
ourselves to only flavour-diagonal currents. A convenient 
parametrization of such operators, for a single SM fermion $\psi$,
is given by

\begin{equation}
\begin{array}{lclclcl}
 \mathcal{O}_{VV}&=& \dis \bar{\chi} \gamma_{\mu}\chi \; 
                        \bar{\psi}\gamma^{\mu}\psi &
                  \qquad \quad&  
 \mathcal{O}_{AA}&= & \dis \bar{\chi} \gamma_{\mu}\gamma_{5}\chi \;
                  \bar{\psi}\gamma^{\mu}\gamma^{5}\psi\\[1ex] 
 \mathcal{O}_{SS}&=& \dis \bar{\chi} \chi \; \bar{\psi}\psi & &
 \mathcal{O}_{PP}&=& \dis \bar{\chi} \gamma_{5}\chi \;\; 
                        \bar{\psi} \gamma_{5} \psi\\[1ex]  
 \mathcal{O}_{VA}&=& \dis \bar{\chi} \gamma_{\mu}\chi \;\;
                          \bar{\psi} \gamma^{\mu} \gamma^{5}\psi & &
 \mathcal{O}_{AV}&=& \dis \bar{\chi} \gamma_{\mu}\gamma_{5}\chi \;\;
                        \bar{\psi} \gamma^{\mu}\psi\\[1ex]
 \mathcal{O}_{SP}&=& \dis \bar{\chi} \chi \;\; \bar{\psi} i \gamma_{5} \psi & &
\mathcal{O}_{PS}&=& \dis \bar{\chi} i \gamma_{5}\chi \; \;
                    \bar{\psi} \psi ,
\end{array}
\label{eq:oper}
\end{equation}
with the subscripts on the operators reflecting the Lore-ntz structure.
The full interaction Lagrangian  involving $\chi$ 
can, then, be parametrized as
\begin{equation}
{\cal L}_{\rm int} = \sum_f \, \sum_{MN \in \rm Operators} \, 
                   \frac{g^f_{MN}}{\Lambda^2} \, {\cal O}_{MN}^f , 
\end{equation}
where $ {\cal O}_{MN}^f \equiv (\bar\chi\Gamma_M\chi) \, 
(\bar f\Gamma_N f) $ is a typical operator amongst those listed in
eq.(\ref{eq:oper}) and $g_{MN}^f$ the corresponding
strength. $\Lambda$ refers to the cut-off scale of the effective
theory.

\par Before we analyze any further, it is useful to consider 
the DM pair-annihilation  cross section engendered by these 
operators~\cite{Zheng:2010js}. Restricting ourselves to a single 
species\footnote{For multiple fermions, the cross-sections, of course, 
add incoherently.} $f$, and denoting $x_i \equiv 2 m_i^2 / s$, we have
\begin{subequations}
\begin{eqnarray}
\sigma^{\rm ann}_{VV} & = & 
      \frac{\left(g^f_{VV}\right)^2 \, N_c}{16 \, \pi \, \Lambda^4} \, 
          \frac{\beta_f}{\beta_\chi} \, s \, 
\left( 1+ x_\chi \right)  \,
\left( 1+ x_f \right) \label{sigma_ann_VV}\\
 \sigma^{\rm ann}_{AA} & = & 
    \frac{\left(g^f_{AA}\right)^2 \, N_c}{12 \, \pi \, \Lambda^4} \, 
      \frac{\beta_f}{\beta_\chi} \, s \, 
\left [ 1 - 2 \, (x_\chi + x_f) + 7  x_\chi  x_f\right]\label{sigma_ann_AA}\\
 \sigma^{\rm ann}_{VA} & = & 
 \frac{\left(g^f_{VA}\right)^2 \, N_c}{12 \, \pi \, \Lambda^4} \, 
        \frac{\beta_f^3}{\beta_\chi} \, s \, 
	\left( 1 + x_\chi\right) \label{sigma_ann_VA}\\
 \sigma^{\rm ann}_{AV} & = &  
 \frac{\left(g^f_{AV}\right)^2 \, N_c}{12 \, \pi \, \Lambda^4} \, 
\beta_f \, \beta_\chi \, s \, \left(1 + x_f\right ) \label{sigma_ann_AV}\\
 \sigma^{\rm ann}_{SS} & = &  
 \frac{\left(g^f_{SS}\right)^2 \, N_c}{16 \, \pi \, \Lambda^4} 
 \, \beta_f^3 \, \beta_\chi \, s
\label{sigma_ann_SS}\\
 \sigma^{\rm ann}_{PP} & = &  
 \frac{\left(g^f_{PP}\right)^2 \, N_c}{16 \, \pi \, \Lambda^4} \, 
         \frac{\beta_f}{\beta_\chi} \, s 
\label{sigma_ann_PP}\\
 \sigma^{\rm ann}_{SP} & = &  
 \frac{\left(g^f_{SP}\right)^2 \, N_c}{16 \, \pi \, \Lambda^4} \, 
\beta_f \, \beta_\chi \, s
\label{sigma_ann_SP}\\
 \sigma^{\rm ann}_{PS} & = &  
 \frac{\left(g^f_{PS}\right)^2 \, N_c}{16 \, \pi \, \Lambda^4} \, 
 \frac{\beta_f^3}{\beta_\chi} \, s \ ,\label{sigma_ann_PS}
\end{eqnarray}
\end{subequations}
where $N_c = 1 \, (3)$ for leptons (quarks) and $\beta_{\chi, f}$ are
the center-of-mass frame velocities of $\chi$ ($f$) with  $\beta_i = \sqrt{1 - 2 \, x_i}$.
Clearly, for non-relativistic DM particles, $ \sigma^{\rm ann}_{SS}$
and $ \sigma^{\rm ann}_{SP}$ are the smallest. Similarly, for 
$m_{\chi} \gg m_f$ (as is the case except when $f$ is the 
top-quark), we have
\[
\sigma^{\rm ann}_{VA} \approx  \sigma^{\rm ann}_{VV} \ , 
\qquad 
\sigma^{\rm ann}_{AV} \approx   \sigma^{\rm ann}_{AA} \ ,
\qquad 
\sigma^{\rm ann}_{PP} \approx   \sigma^{\rm ann}_{PS} \ .
\]
These approximations would prove to be of considerable 
use in further analysis.

Dark matter interactions, as exemplified by equation \eqref{eq:oper}, have been 
well studied in the context of the LHC~\cite{Bhattacherjee:2012ch,Goodman:2010ku,
Carpenter:2012rg,Petrov:2013nia,Carpenter:2013xra,Berlin:2014cfa,Lin:2013sca,
Bai:2010hh,Birkedal:2004xn,Gershtein:2008bf,Fox:2011fx,Bai:2012xg,Crivellin:2015wva,Petriello:2008pu}. Clearly, the LHC would be more sensitive to operators
involving the quarks. In this paper, we would be primarily interested in the orthogonal
scenario, namely one wherein the DM current couples primarily to the SM leptons.
To this end, we  will assume that $g^q_{MN} = 0$ for all of the operators.
As we shall see in the next section, this would render insensitive
even the dedicated direct search experiments.

 \section{Relic Density of Leptophilic Dark Matter}
\label{sec:relicdensity}
Before we proceed further with our analysis, it behooves us to consider
the existing constraints on such a DM candidate. This is best
understood if we consider a single operator at a time, and, henceforth, 
we shall assume this to be so. In other words, all operators bar the
one under discussion shall be switched off. As to the couplings
$g^f_{MN}$, we consider two cases\footnote{Note that such a
normalization of couplings is quite standard in effective field
theories.}, namely (a) Leptophilic: $g^l_{MN} = 1$ for all leptons $l$,
and (b) Electrophilic: $g^e_{MN} = 1$ with all others vanishing.

Concentrating first on the epoch where the DM $(\chi)$ has frozen out, 
let us consider their pair-annihilation to produce two 
light particles $(\ell)$, which, we assume, is in complete 
equilibrium, viz. $ n_\ell = n_{\ell_{eq}} $. 
The Boltzmann equation gives 
us:
\begin{equation}
  a^{-3} \, \frac{d}{dt} (n_{\chi}a^{3}) = \langle\sigma v\rangle 
  \left[n^2_{\chi_{eq}}-n_{\chi}^{2} \right]
\label{eqn:boltzeqn1}
\end{equation}
where $a\equiv a(t)$ is the scale factor and $\langle \cdots\rangle$
represents the thermal average. Here, $v$ is the relative velocity of
a pair of $\chi$'s, and $\sigma$ is the annihilation cross section.

In the radiation era, which we are interested in, the temperature
scales as $a^{-1}$.  We can, therefore, rewrite the Boltzmann equation, 
in terms of new dimensionless variables
\begin{equation}
Y \equiv \frac{n_{\chi}}{T^3} \qquad \mbox{and} \qquad
x \equiv \frac{m_{\chi}}{T} \ ,
\end{equation}
as 
\begin{equation}
\frac{dY}{dx} = -\frac{m^3_{\chi} \, \langle\sigma \,v\rangle}{H(m_{\chi}) \, x^{2}}\,
                 \left( Y^{2} - Y_{eq}^{2} \right) \ ,
\label{eqn:boltzeqna}
\end{equation}
where $H(m_{\chi})$ is the Hubble rate when the temperature equals $m_{\chi}$ and is given by 
\begin{equation}
H(m_{\chi})= \sqrt{\frac{4 \, \pi^3 \, G_{N} \, g_{\star}}{45}} \, m^2_{\chi}. 
\end{equation}
Here  $ g_{\star}$ is the effective number of the degrees of freedom at that epoch.

\par To find the present density of DM particles, we need to find the
solution of eq.~\eqref{eqn:boltzeqna} in terms of the final freeze out
abundance, $Y_{\infty}$ (at $ x = \infty$). While this, can be done only 
numerically, it is instructive to consider an approximate analytic solution.
At early times ($T \gg m_\chi$) reactions proceeded rapidly, and $n_{\chi}$
was close to its equilibrium value, $n_{eq} \propto T^3$ , or $Y \approx Y_{eq}$. 
As the temperature falls below $m_{\chi}$, the equilibrium abundance 
$Y_{eq}= g/(2 \, \pi)^3 \, x^{3/2} \, e^{-x}$ is exponentially suppressed.
Conseque-ntly, the DM particles become so rare that they do not find each
other fast enough to maintain the equilibrium abundance. This is the onset
of freeze-out. In other words, well after freeze out, $Y \gg Y_{eq}$ and 
\begin{eqnarray}
\frac{dY}{dx} = -\frac{m^3_{\chi} \, \langle\sigma v\rangle}{H(m_{\chi}) \, x^{2}} \,
                  Y^{2}
\label{eqn:boltzeqnb}
\end{eqnarray}
Integrating equation ~\eqref{eqn:boltzeqnb} from the freeze out temperature
$x_{f}$ until very late times $x = \infty$,
we get 
\begin{eqnarray}
Y_{\infty} \simeq \frac{x_f \, H(m_{\chi})}{m^3_{\chi} \langle\sigma v\rangle} .
\end{eqnarray}
\begin{figure*}[htb]
  \centering
  \begin{tabular}{cc}
  \includegraphics[width=0.5\textwidth,clip]{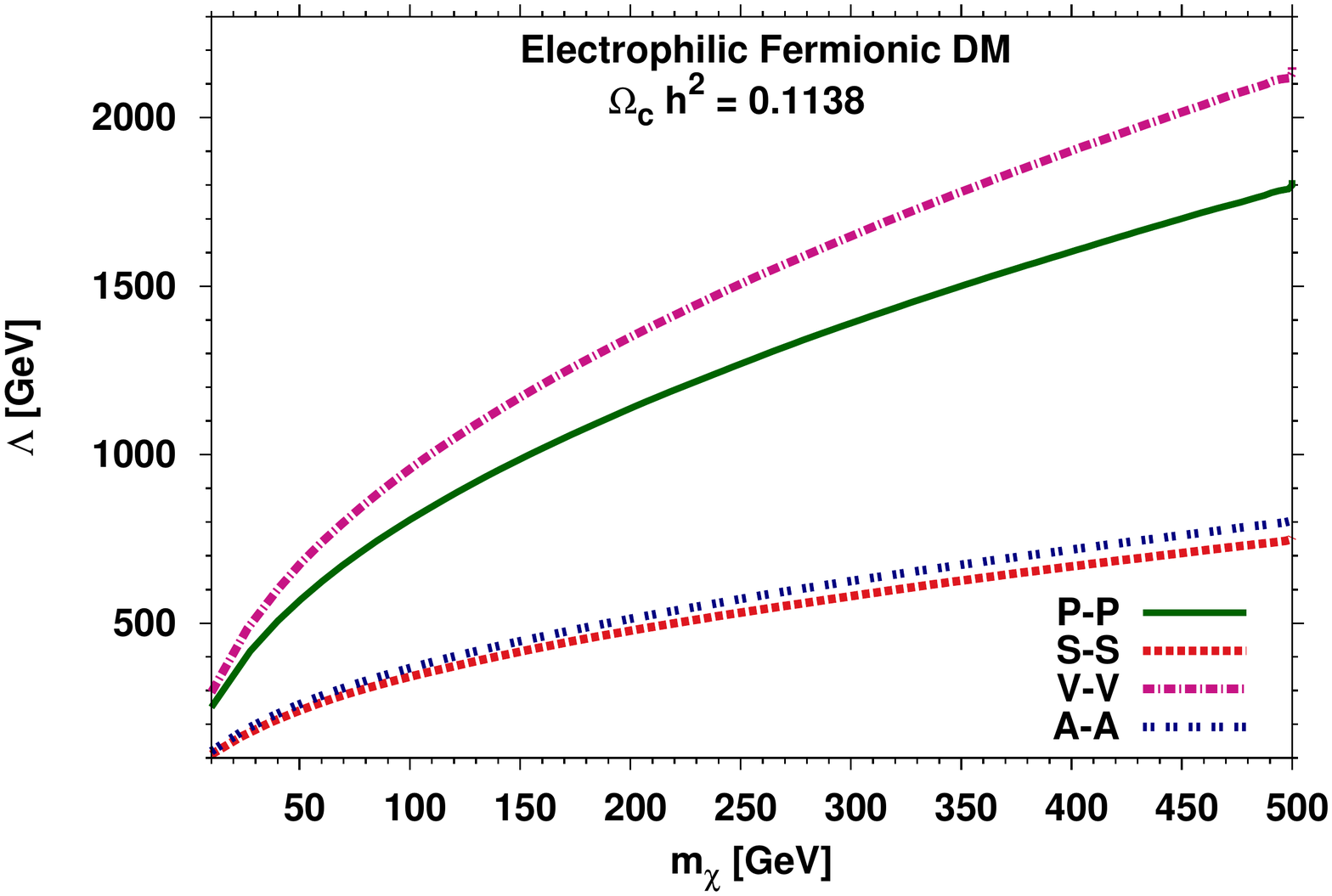} &
  \includegraphics[width=0.5\textwidth,clip]{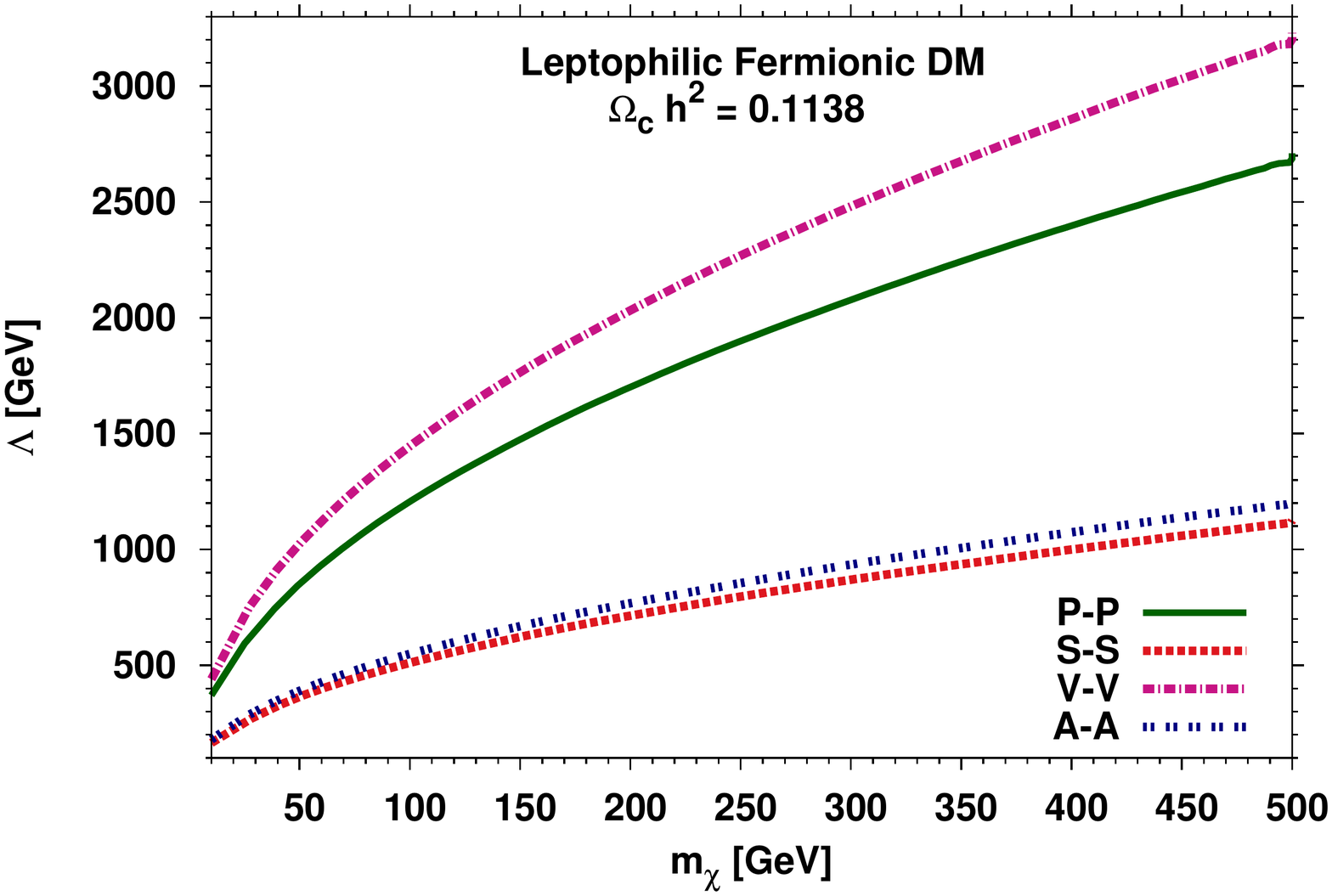}
  \end{tabular}
  \caption{\small \em{Contours in the $\Lambda$--$m_\chi$ plane based
      on the cold dark matter density $\Omega_c h^2 = 0.1138 \pm
      0.0045$ obtained from the 9-year WMAP data
      \cite{Bennett:2012zja}. These curves are Obtained using
      $micrOmegas \mbox{-} v \mbox{-}3.5.5$~\cite{Belanger:2001fz},
      the regions below the curves are allowed by the relic density bounds.}}
\label{fig:relicdensity} 
\end{figure*}

\par The energy density, for the now non-relativistic DM,  
$\rho_{\chi} = m_{\chi} \, n_{\chi}$, 
falls as $a^{-3}$. The number density $n(a(T_i),T_i)$ in this post freeze
out epoch at any given  temperature $T_i$ behaves as $Y_{\infty} \, T_i^3$.
The number of the degrees of freedom also changes with the temperature from
$g_f(x_f)$ at the freezing epoch  to that of today $g_0(x_0)\sim$ 3.36 at 
the present day temperature $T_0$. The present energy density of DM can,
thus, be re-written as
\begin{eqnarray}
 \rho_\chi=m_\chi\, Y_\infty\,T_0^3\,\left(\frac{g_0(x_0)}{g_f(x_f)}\right).
 \end{eqnarray}  It is customary to parametrize
$\rho_{\chi} \equiv \Omega_{\chi} h^2 \, \rho_c $, 
where $\rho_{c}=1.05375 \times 10^{-5} h^2 \left( GeV/c^2 \right)$ cm$^{-3}$
is the critical density of the universe and the Hubble constant today being expressed 
as $H_0 = h \times 100$ \, km \, s$^{-1}$ \, Mpc$^{-1}$. We have, then,
\begin{equation}
\Omega_{\chi} h^2 = \sqrt{\frac{4 \, \pi^3 G_N g_{\star}(x_f)}{45}} \; 
\frac{x_f \, T^3_0 \,  g_0}{\rho_c \, \langle\sigma \, v\rangle \, g_{\star}(x_f)}
\ .
\end{equation}
Using $ T_0$ = 2.72548 $\pm$ 0.00057 K \cite{Fixsen:2009ug}, the contribution of
the Dirac fermion $\chi$, with its two degrees of freedom, is 
\begin{eqnarray}
\Omega^{(\chi)}_{DM} h^2 \approx 2 \times x_f \,
\frac{1.04 \times 10^9 {\rm GeV}^{-1} }{m_{pl} \, \sqrt{g_{\star}(x_f)} \, \langle\sigma v\rangle} \ .
\label{eqn:relic}
\end{eqnarray}
\par While the WMAP 9-year data gives $\Omega_{DM} h^2$ = 0.1138 $\pm$ 
  0.0045, the Planck 2015 results\cite{Ade:2015xua} suggest a
  marginally different value, namely, $\Omega_{DM} h^2$ = 0.1199 $\pm$ 
  0.0022. This translates to a
  strict relation in the mass-coupling plane for $\chi$. However, a
  conservative assumption would be that the relic density of the
  $\chi$ does not saturate the observed DM energy density, or
  $\Omega_\chi < \Omega_{DM}$. This would impose a lower bound on the
  self-annihilation cross-section for the $\chi$, or, in other words,
  an upper bound on $\Lambda$.

\subsection{The annihilation cross-sections}
In the calculation of the probability of a DM particle being 
annihilated by another one, it is useful to consider the rest 
frame of the first and re-express the aforementioned 
annihilation cross sections in terms of the velocity $v$ 
of the second particle in his frame. Clearly,
$v = \left(2 \,\beta_\chi\right)/ \left(1 + 2 \,\beta_\chi^2\right)$ 
and, working with a small $v$ expansion (relevant for  non-relativistic DM), we have
$\beta_\chi \simeq \frac{v}{2} \, \left[1 + \frac{v^2}{4} + \frac{v^4}{8}
\right] + \mathcal{O}(v^7)$ and $ s \simeq m_\chi^2 \, \left[4 + v^2 + 
\frac{3}{4} v^4\right]$  $+ \mathcal{O}(v^6) $. The corresponding 
$v$-expansions, to ${\cal O}(v^2)$, are
\begin{subequations}
\begin{eqnarray}
\sigma^{\rm ann}_{SS} \, v &\simeq& \dis
     \frac{\left(g^l_{SS}\right)^2 \,N_c}{8 \, \pi\, \Lambda^4} \,
         \left(1-\xi_f \right )^{3/2} \, m^2_{\chi}v^2 \label{eqn:sigmav_SS}\\
 \sigma^{\rm ann}_{PP} \, v &\simeq& \dis
     \frac{\left(g^l_{PP}\right)^2 \,N_c}{2 \, \pi \, \Lambda^4} \,
          \sqrt{1-\xi_f} m^2_{\chi} \,
\left [ 1 + \frac{\xi_f \, v^2}{8 \, \left ( 1-\xi_f \right )} \right ] \, \label{eqn:sigmav_PP}\\
 \sigma^{\rm ann}_{VV} \, v &\simeq& \dis
     \frac{\left(g^l_{VV}\right)^2 \,N_c}{2 \, \pi \, \Lambda^4} \,
          \sqrt{1-\xi_f} \, \left ( 2 \, m^2_{\chi} + m^2_f  \right ) \,\nn\\
&&\left [ 1 + \frac{-4 + 2 \,\xi_f + 11 \, \xi_f^2}{24 \, 
\left ( 1-\xi_f \right ) \, \left ( 2+\xi_f \right )} v^2\right ] \label{eqn:sigmav_VV}\\
 \sigma^{\rm ann}_{AA} \, v &\simeq& \dis 
      \frac{\left(g^l_{AA}\right)^2 \,N_c}{2 \,\pi\, \Lambda^4} \,
          \sqrt{1-\xi_f} \, m^2_f \,\nn\\
&&\left [ 1 + \frac{ 8 \, \xi_f^{-1} - 28 + 23 \, 
\xi_f}{24 \, \left ( 1-\xi_f \right )} v^2\right ] \ ,\label{eqn:sigmav_AA}
\end{eqnarray}
\end{subequations}
where $\xi_f \equiv m_f^2 / m_\chi^2$. Clearly, the $v$-independent terms 
emanate from $s$-wave scattering alone, while the 
${\cal O}(v^2)$ piece receives contributions  from both $s$-wave 
and $p$-wave annihilation processes.

\par The quantity of interest is not just $\sigma^{\rm ann}$, but the 
thermal average $\langle \sigma^{\rm ann}_{MN} v \rangle$ since this 
provides a measure of the rate at which a DM particle will meet 
another with the appropriate velocity and get annihilated. 
 Assuming $\chi \chi \rightarrow e^+ e^−$ to be the dominant channel into
 which the $\chi$'s annihilate, we can estimate the thermally-averaged value of
 $ \sigma(\chi \chi \rightarrow e^+ e^-) \,v$.

The inferred value of $\Omega_{\chi} h^2$ when ascribed to WIMPs with a 
mass of a few hundred GeVs, requires $ \langle \sigma(\chi \chi \rightarrow e^+ e^-) \,
v\rangle \approx 3 \times 10^{-26} \, cm^3 \, s^{-1}$. Taking cue from this,
we analyze upper bounds on $ \langle \sigma(\chi \chi \rightarrow e^+ e^-) \,
v\rangle $ in section ~\ref{sec:processes} from the sensitivity study 
of the DM signatures from ILC for the kinematically accessible DM mass range.

\par In Figure~\ref{fig:relicdensity} we present the relic density plots 
for the interactions parametrized by the operators mentioned in 
section~\ref{sec:fermionicdm} corresponding to the Leptophilic and 
Electrophilic scenarios. We implement the model containing effective interactions
of DM particle with the SM sector using FeynRules ~\cite{Christensen:2008py} and 
generate the model files for CalcHEP which is required for the relic density calculation
in micrOmegas~\cite{Belanger:2001fz}. These results have been verified with
MadDM~\cite{Backovic:2013dpa,Backovic:2015tpt}.

The curves in Figure~\ref{fig:relicdensity} imply that the correct amount of energy
density requires the mass of dark matter particles to increase with $\Lambda$. 
We can infer this behaviour from~\eqref{eqn:relic} as
$\Omega_{\chi} \, h^2$ mainly depends on $\langle\sigma
v\rangle$. From equations~\eqref{eqn:sigmav_SS}-\eqref{eqn:sigmav_AA}
we see that the thermal average in equation
~\eqref{eqn:relic} i.e. $\left \langle \sigma v \right \rangle$ is
proportional to $m_{\chi}^2/\Lambda^4$ for $m_{\chi}$ values greater
than $\sim$ 10 \, GeV. This implies that for a fixed value of
$\Omega_{\chi} \, h^2$, $\Lambda^4/m_{\chi}^2$ is a constant. This, in turn,
translates to $\Lambda \propto \sqrt{m_{\chi}}$.

\section{Direct and Indirect Detection }
\subsection{DM-electron scattering:}
We now turn to direct detection search, where both electrophilic and
leptophilic DM are scattered by \begin{enumerate}
\item[(a)] the bound electron of an atom and hence the recoiled
  electron is ejected out of the atom,
\item[(b)] the bound electron and the recoil is taken by the atom as a
  whole and
\item[(c)] the quarks, where the DM is attached to the loop of charged
  virtual leptons which in turn interacts with the quarks {\it via}
  photon exchange and hence the nucleus recoil is measured in the
  detector.
\end{enumerate}
\par
\begin{figure}[h]
\centering
\includegraphics[width=0.5\textwidth,clip]{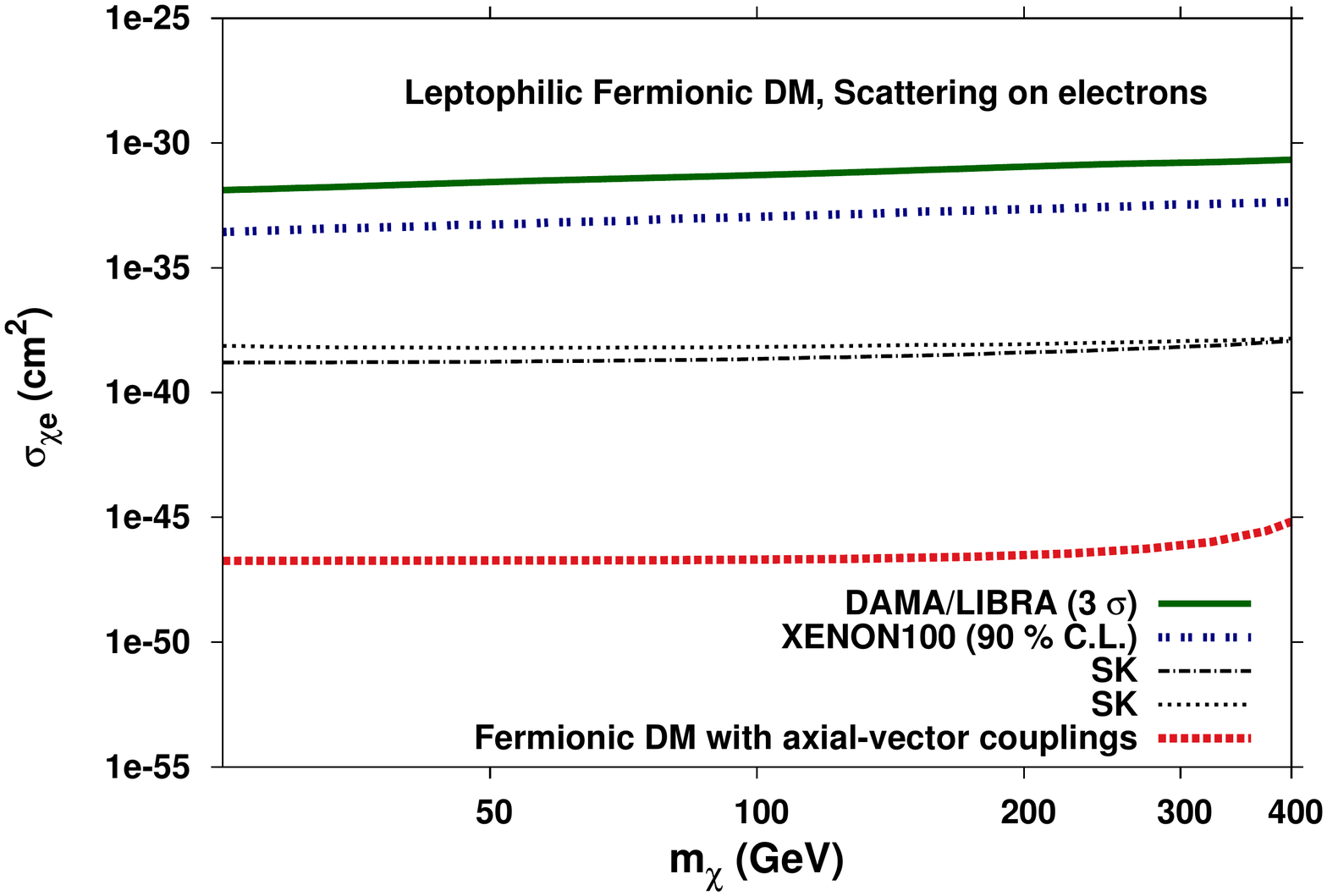}
\caption{\small \em{ Scattering cross-section of DM on free electron with 
         the axial-vector interactions $\sigma\left( \chi\, e^-\to \chi\, e^-\right)$ 
         is depicted for the DM mass varying between 25-400 GeV. Exclusion 
         plots from DAMA at 90\% and 3$\sigma$ C.L. for the case of DM-electron 
         scattering are also shown \cite{Kopp:2009et}. Bounds at 90\% C.L. are 
		 shown for XENON100 from inelastic DM-atom scattering \cite{Aprile:2015ade}. 
		 The dashed curves show the 90\% CL constraint from the Super-Kamiokande
         limit on neutrinos from the Sun, by assuming annihilation into $\tau^+\tau^-$
         or $\nu\bar\nu$ \cite{Kopp:2009et}.}}\label{directdetection}
\end{figure}

 In this article we restrict our study for the DM-electron scattering only. We analyze the free electron scattering cross-sections with the
DM for the scalar, pseudoscalar, vector and axial couplings
respectively and are given as
\begin{subequations}
\begin{eqnarray}
 \sigma^{SS}_{\chi e} &=& \dis
     \frac{\left(g^l_{SS}\right)^2 \, m^2_e}{\pi\, \Lambda^4} \label{eqn:sigmachie_SS}\\
 \sigma^{PP}_{\chi e} &=& \dis
     \frac{\left(g^l_{PP}\right)^2 \, m^2_e}{3 \, \pi \, \Lambda^4} \frac{m^2_e}{m^2_{\chi}} \, v^4 \label{eqn:sigmachie_PP}\\
 \sigma^{VV}_{\chi e} &=& \dis
     \frac{\left(g^l_{VV}\right)^2 \, m^2_e}{\pi \, \Lambda^4} \label{eqn:sigmachie_VV}\\
 \sigma^{AA}_{\chi e} &=& \dis 
      \frac{3  \left(g^l_{AA}\right)^2 m^2_e}{\pi\, \Lambda^4}\label{eqn:sigmachie_AA}
\end{eqnarray}
\end{subequations}
where $v\sim 10^{-3} c$, is the DM velocity in our halo. We observe
that the cross section with the axial-vector couplings of DM dominates
over all others. The pesudoscalar type coupling is highly suppressed
due to its dependence on the fourth power of the electron mass and velocity.
Respective cross sections for the case of bound electrons are likely
to be enhanced than that of free electrons nearly by a factor of
$\sim 10^5$ \cite{Kopp:2009et}. Using the constraints from the relic
density on the lower bound of the couplings for a given DM mass
$m_\chi$, in Figure ~\ref{directdetection}, we present the
scattering cross section of the DM with free electrons {\it via}
the most dominant channel with axial-vector couplings. We compare our
result with the recent $3\sigma$ limits from the DAMA/ LIBRA
experiment \cite{Kopp:2009et} and 90\% C.L. data from XENON100
\cite{Aprile:2015ade}. However, even with the inclusion of the
correction factor arising from the the bounded electrons, the
contribution of the axial-vector couplings of DM to the scattering
cross-section are much below these experimental upper bounds.

\begin{figure}[h]
\centering
  \includegraphics[width=0.5\textwidth,clip]{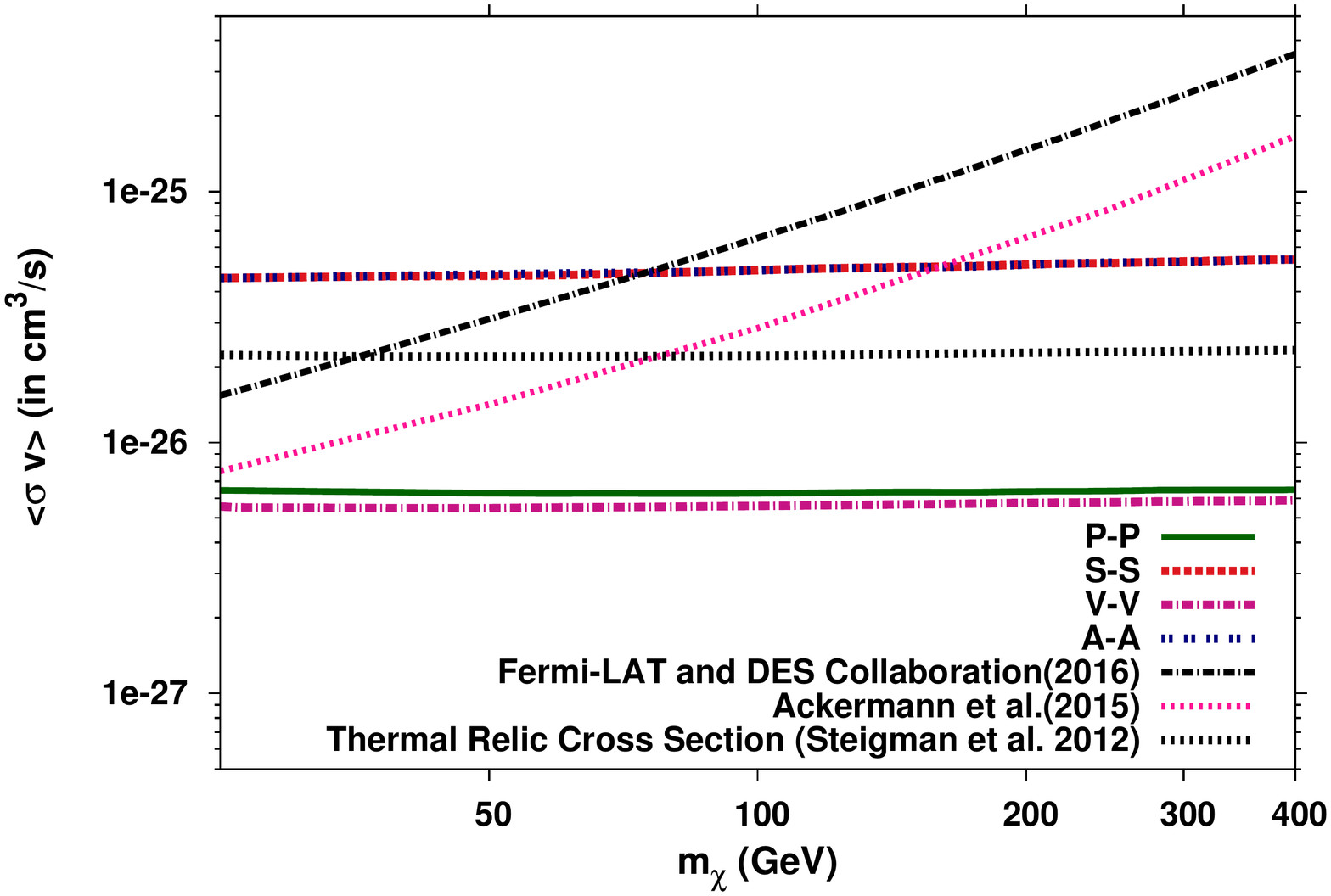} 
  \caption{\small \em{Thermally averaged DM annihilation
      cross-sections versus the DM mass $m_\chi $ using the relic
      density allowed lower bound on S-S, P-P, V-V and A-A couplings
      for DM. These cross-sections are compared with the median of the
      DM annihilation cross section derived from a combined analysis
      of the nominal target sample for the $\tau^+\tau^-$ channels
      \cite{Fermi-LAT:2016uux}.}}
\label{fig:indrdetectn} 
\end{figure}

 \par Indeed, the {\sc Xenon100} collaboration
has constrained the $\sigma(\chi e \to \chi e) \lsim 10$ pb at 90\%
C.L. for\footnote{The exclusion limits do have a dependence on
  $m_\chi$ with the maximum sensitivity occurring at $m_\chi \sim 2
  \gev$ and relaxing by an order of magnitude for the highest $m_\chi$
  values.}  $0.6 \gev < m_\chi < 1 \tev$ ~\cite{Aprile:2015ade} for
the case of axial-vector coupling. Expectedly, these constraints are
much stronger (by nearly an order of magnitude) than the cross
sections required to explain the {\sc DAMA/Libra}
results~\cite{Bernabei:2013xsa}. Interestingly, the strongest bounds
emanate from solar physics. In general, the neutrino flux coming from
DM annihilations inside the sun is proportional to the DM scattering
cross section. But, working in the region where DM capture and its
annihilations inside the star are in equilibrium, makes the neutrino
flux independent of the DM annihilation. As the DM particles can be
trapped in the solar gravitational field, their annihilation into
neutrinos would modify the solar neutrino spectrum and, consequently,
the SuperKamiokande measurements ~\cite{Kopp:2009et}. This translates
to $\sigma(\chi \bar\chi \to \tau^+ \tau^-) \lsim 0.1$ pb and
$\sigma(\chi \bar\chi \to \nu \bar\nu) \lsim 0.05$ pb which are
relevant for the general leptophilic case with axial-vector
coupling. Understandably, no such bound exists for the purely
electrophilic case.
\subsection{DM Annihilation:}
Next we discuss the bounds from the indirect detection experiments,
wherein we can directly use the cross sections given in equations
\eqref{sigma_ann_VV}- \eqref{sigma_ann_PS} and fold them with the
local velocity of the DM particle. The latter, of course, would depend
crucially on the particular profile of the DM distribution that one
adopts. In~\cite{Fermi-LAT:2016uux}, the expected sensitivity is given
as a limit on the thermally averaged DM annihilation cross section for
the $\tau^+ \, \tau^-$ channel.

From equation~\eqref{eqn:sigmav_SS}-\eqref{eqn:sigmav_AA}, the
thermally averaged annihilation cross- section of a leptophilic DM of
mass $m_\chi$ can be estimated using the relic-density allowed lower
bound on the scalar, pesudoscalar, vector and axial-vector couplings
of DM.  The annihilation cross-section for the leptophilic DM is
marginally higher than electrophilic DM and roughly are of the same
order of magnitude for $m_\chi \sim 10-400$ GeV.  We compute the
annihilation cross-section for the leptophilic DM with these lower
bounds on DM couplings and depict in Figure ~\ref{fig:indrdetectn}. We
compare our thermally averaged cross-section of leptophilic DM with
the bounds from the current indirect detection searches
\cite{Fermi-LAT:2016uux}.

\section{Interesting Signatures at ILC}
\label{sec:processes}
\begin{table*}[!h]\footnotesize
\centering
\begin{tabular*}{\textwidth}{c|@{\extracolsep{\fill}} cccccc|}\hline\hline
                  &\textit{ILC-500}&\textit{ILC-500P}&\textit{ILC-1000}&\textit{ILC-1000P}\\
$\sqrt{s} \left( \textit{in GeV}\right )$& $500$ & $500$ & $1000$ & $1000$ \\
$L_{int} \left( \textit{in $fb^{-1}$}\right )$& $500$ & $250$ & $1000$ & $500$ \\
$P_{-}$        & 0 & 80\% & 0 & 80\%\\
$P_{+}$        & 0 & 30\% & 0 & 30\%\\\hline\hline
\end{tabular*}
\caption{\small \em{Accelerator parameters for each of the run scenarios considered in this paper.
                    $P_{-}$ and $P_{+}$ represent the electron and positron polarizations respectively.}}
\label{table:accelparam}
\end{table*}
\begin{table*}[h!]\footnotesize
\begin{center}
\begin{tabular}{c||c|c||c|c|}
\hline\hline
&\multicolumn{2}{c||}{\bf \underline{Unpolarized}}&\multicolumn{2}{c}{\bf \underline{Polarized}}\\

Interactions    &{S-S}&{V-V}&{S-S}&{V-V}\\
$\sqrt{s}$ in TeV   &1.0&1.0 &1.0&1.0\\
${\cal L}$ in fb$^{-1}$    &1000&1000&500 &500\\
$\left(P_{e^-},\, P_{e^+}\right)$    &(0,\,0) &(0,\,0) &(.8,\,-.3)&(.8,\,-.3)\\
\hline
\underline{\bf DM Mass in GeV}&&&&\\
 75  &9.46&3.27 &30.78&15.79 \\ 
 225 &3.76&2.93 &14.34&14.31\\ 
 325 &0.72&1.94 &3.00 &9.87\\  
&&&&\\ \hline       
\end{tabular}
\end{center}
\caption{\small \em{The efficiency $S/\sqrt{(S+B) + \delta^2_{sys} (S+B)^2}$ 
        with $\delta_{sys} = 1 \%$ for the process $e^+e^-\to j\,j+\not\!\!E_T $ 
        on imposition of the basic cuts for $\Lambda $ = 1 TeV corresponding to 
        the three representative values of the DM mass with unpolarized and 
        polarized initial beams. }}
\label{table:cs}
\end{table*}

\begin{figure*}[tbh]
        \centering
\includegraphics[width=7.2cm,height=6cm]{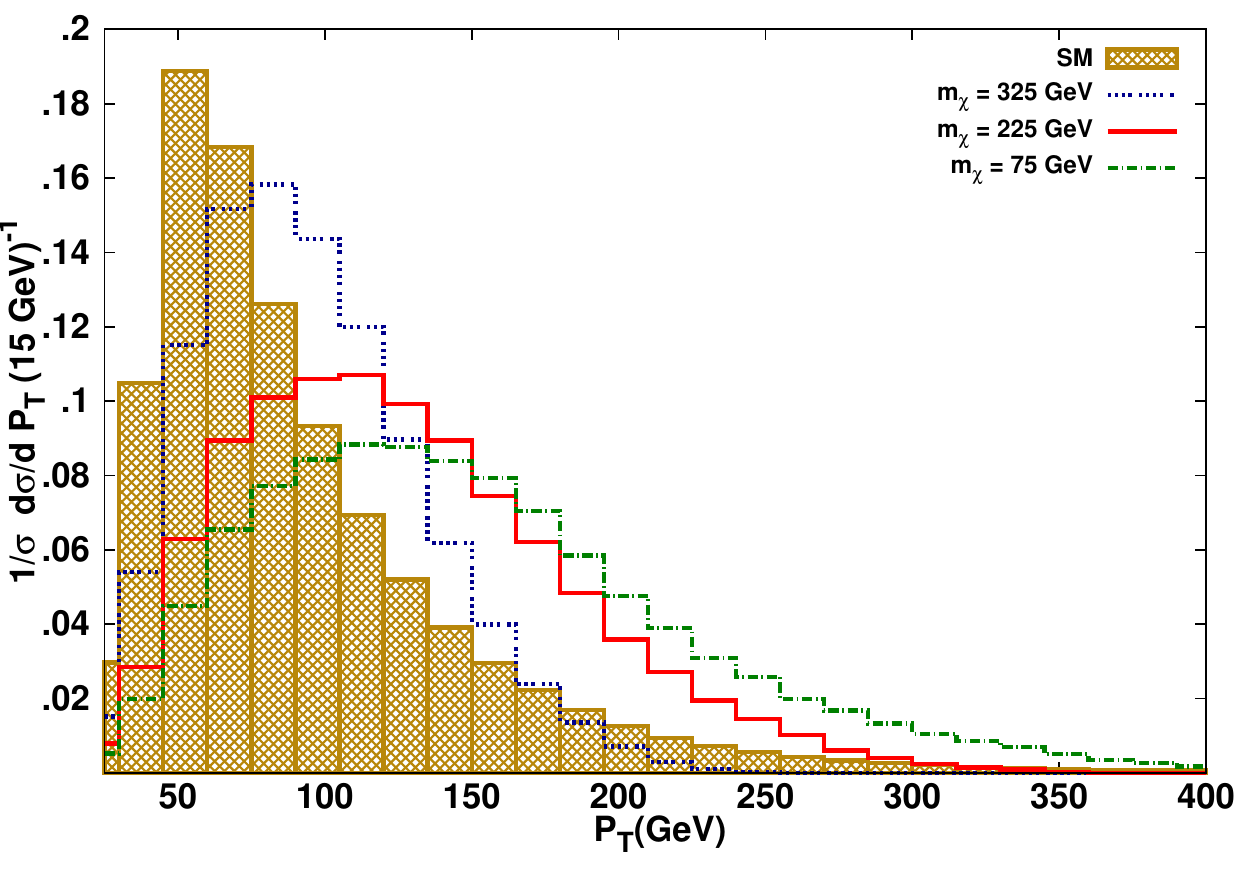}    
\includegraphics[width=7.2cm,height=6cm]{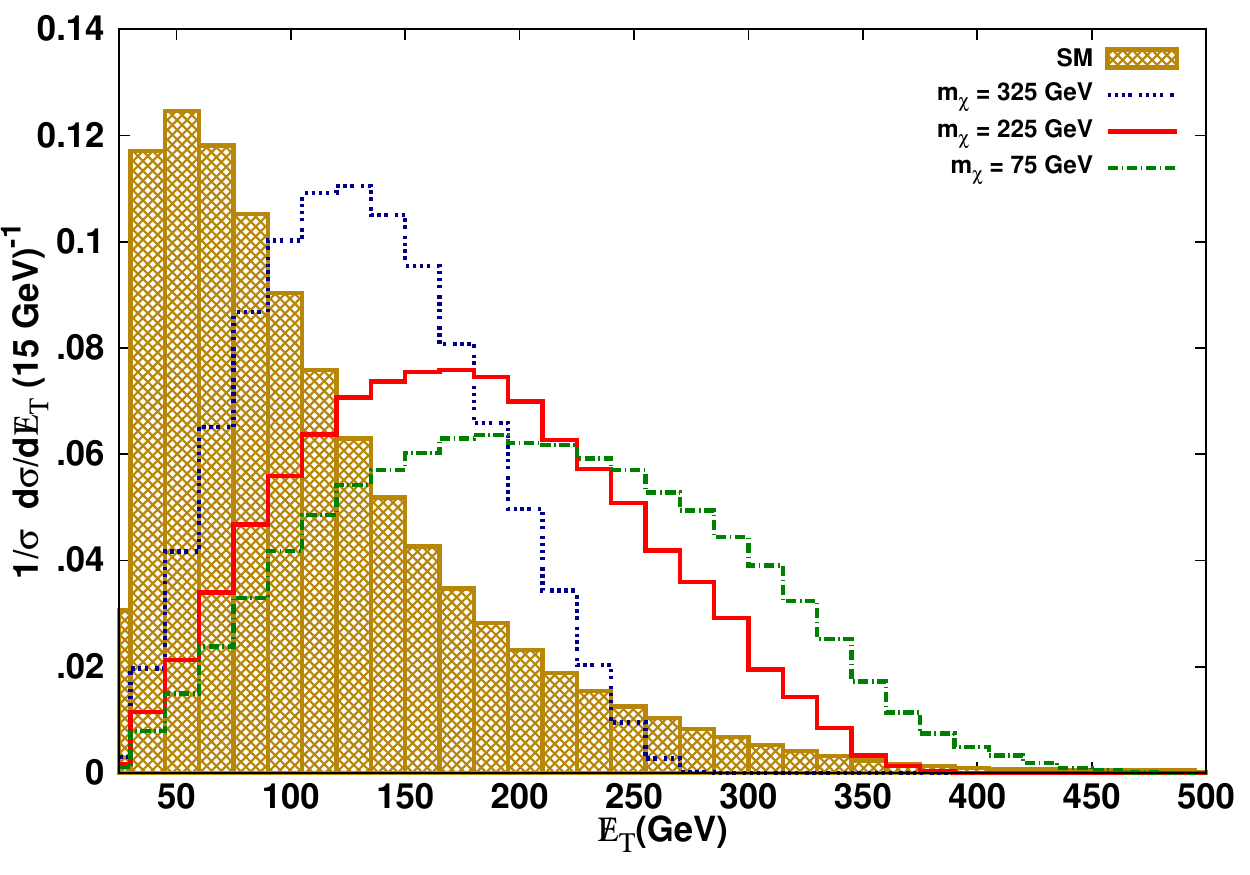}
\includegraphics[width=7.2cm,height=6cm]{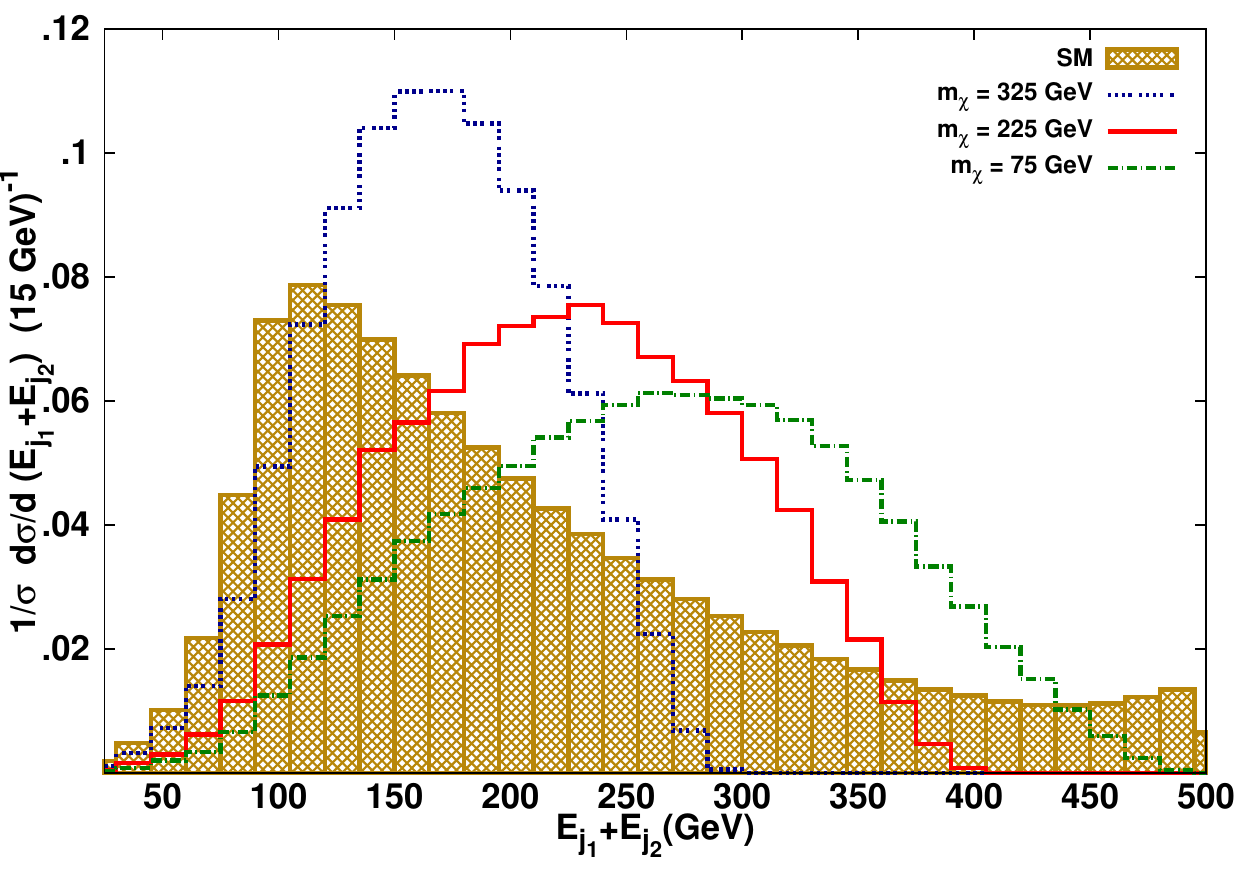}        
\includegraphics[width=7.2cm,height=6cm]{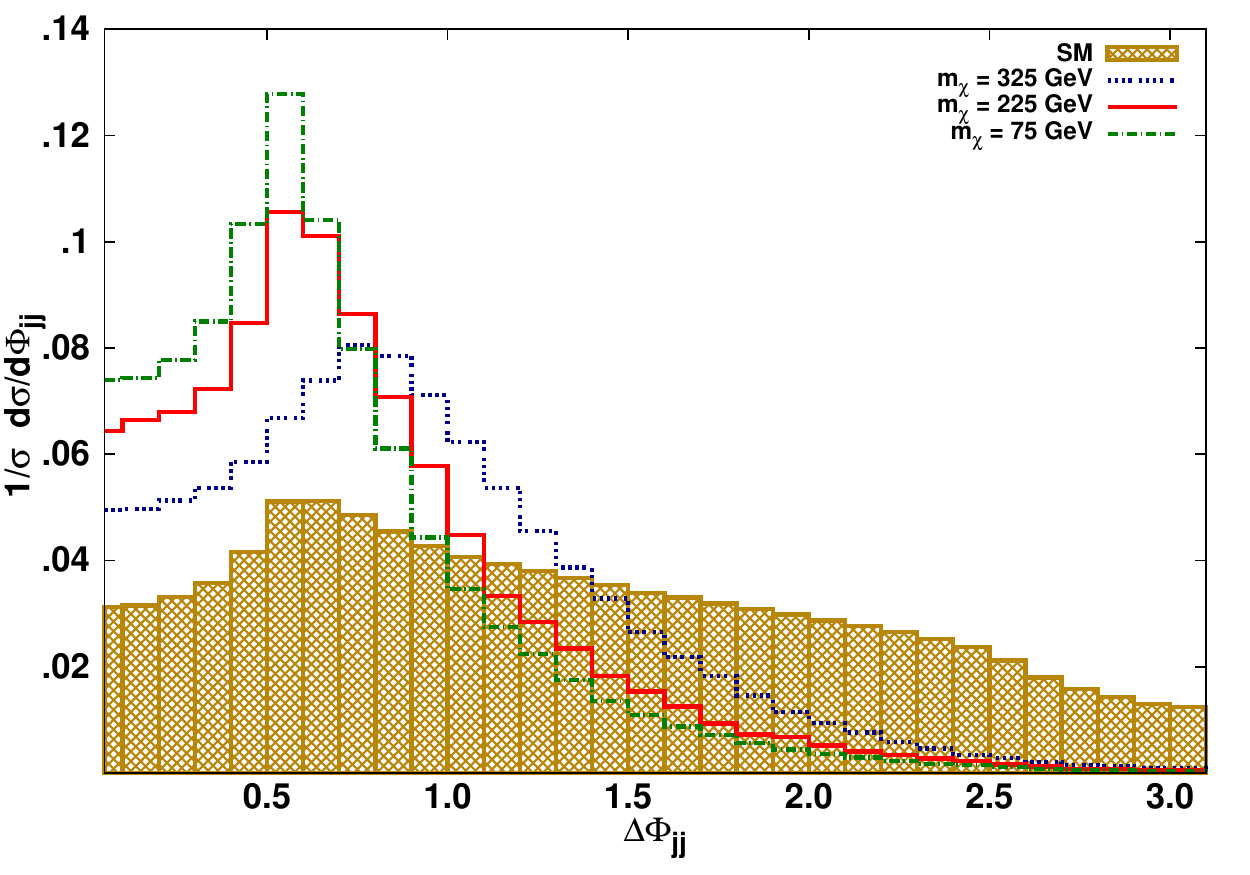}
        \caption{\small \em{Normalized 1-D differential distribution
        of the cross-section {\it w.r.t.} kinematic observable
        ${P_T}_{j_1}$, $\not \!\! E_T$, sum of energy of jet pairs and
        $\Delta\phi_{jj}$ for the background process (shaded
        histograms) and the $Z\to j\, j$ associated DM pair production
        at the three representative values of $M_{\chi}$ = 75, 225 and
        325 GeV with $\sqrt{s}$ = 1 TeV, $\Lambda$ =1 TeV and
        $g_{SS}^l$ ($g_{PP}^l$, $g_{SP}^l$, $g_{PS}^l$) =1. The bin
        width is 0.1 for the $\Delta\phi_{ii}$ distribution while it
        is 15 GeV for the $\not \!\! E_T$ and $E_{j_1}+E_{j_2}$. }}
        \label{fig:distrSpgs}
\end{figure*}
While DM particles can be produced in many different processes
at a given collider, only a few of them are, potentially, of
interest. Remembering that the DM particle has to be produced only in
pairs, and that there must be at least one visible particle in the
final state so as to register the event in a detector, the simplest
process at a linear collider is, of course, $e^+ e^- \to \bar\chi \chi
\gamma$, and it has, rightly, been well studied in many different
contexts~\cite{Chae:2012bq,Dreiner:2012xm,Yu:2013aca}. Although the
signature, namely monophoton with missing energy-momentum, can be
masqueraded by both detector effects as well as a large and irreducible
background emanating from $e^+ e^- \to \bar\nu_i \nu_i \gamma$, the
kinematic profiles are sufficiently different enough to merit the
possibility of a discovery.

Given the exceeding simplicity of the aforementioned channel,
attention has focussed on it almost exclusively. However, it is worthwhile to explore
complementary channels, like, DM pair-production in association with on-shell
$Z$ or off-shell $Z^\star$, 
\begin{eqnarray}
e^+ e^- \to  \chi\bar{\chi} + Z\, \left(Z\to jj, \mu^+ \mu^- , e^+ e^- \right)
\label{eqn:sig}
\end{eqnarray}
where $j$ refers to a jet arising from any of $u,d,s,c,b$.  In this
paper, we shall not attempt to impose any $b$-tagging (or $b$-veto)
requirements, and, hence, the $b$-jet is equivalent to any other light
quark jet\footnote{While similar analyses have been attempted in
    Refs.~\cite{Neng:2014mga,Yu:2014ula}, we differ substantially from
    their conclusions even when we adopt an approach identical to
    theirs. Furthermore, we present, here, a much more refined
    analysis.}. To effect a realistic analysis, we have to include, 
though, off-shell contributions as well and, thus, need to consider 
$e^+ e^- \to f \bar f + \bar \chi \chi$, where $f \equiv \mbox{jet}, e^-, \mu^-$. 
The corresponding irreducible SM backgrounds emanate from\footnote{In all 
our analysis, the dijet (plus missing energy) signal is actually an inclusive 
one, in that we demand a minimum of two jets.}
\begin{subequations}
\begin{eqnarray}
 e^+ e^- &\rightarrow& 2 \, {\rm jets} + \sum_{i} \nu_i\bar{\nu_i}, \label{eqn:bkgd1}\\ 
 e^+ e^- &\rightarrow& \mu^+ \mu^- + \sum_{i} \nu_i\bar{\nu_i}, \label{eqn:bkgd2}\\
 e^+ e^- &\rightarrow& e^+ e^- + \sum_{i} \nu_i\bar{\nu_i}.
\label{eqn:bkgd3}
\end{eqnarray}
\end{subequations}
respectively. The largest contribution to the first mode
in eqn.~\eqref{eqn:bkgd1} accrues from $e^+ e^- \to Z + \gamma^*/Z^*$ 
with $Z \to \nu_i \bar \nu_i$ and $\gamma^*/Z^*$ splitting to two jets
followed by the $W^+W^-$ fusion channel $e^+e^-\to \nu_e \bar\nu_e W^+ 
W^-\to\nu_e \bar\nu_e jj $ and t-channel W exchange diagrams.
Feynman diagrams analogous to both these sets also contribute  
to the two other processes given in~\eqref{eqn:bkgd2} and~\eqref{eqn:bkgd3}
respectively. The relative strengths of these channels depend on a 
multitude of factors, such as the center-of mass energy, beam polarization
(if any) and the kinematical cuts imposed. 

We use MadGraph5 ~\cite{Alwall:2014hca}  to perform the simulations for both
SM background and the signal. We employ PYTHIA6
~\cite{Sjostrand:2006za} to carry out the parton shower and
hadronization. PGS ~\cite{pgs} is used for Fast detector simulation
with anti-$k_T$ algorithm for jet reconstruction. In accordance with
technical design report for ILC detectors ~\cite{Behnke:2013lya}, we
set the energy smearing parameters of the electromagnetic calorimeter
and the hadronic calorimeter in the PGS card as
\begin{subequations}
\begin{eqnarray}
\frac{\Delta E}{E} &=& \frac{17 \% }{\sqrt{E/ \rm GeV}} \oplus 1\% \label{eqn:pgsparam1}\\ 
\frac{\Delta E}{E} &=& \frac{30 \%}{\sqrt{E/ \rm GeV}}
\label{eqn:pgsparam2}
\end{eqnarray} 
\end{subequations}
The other accelerator parameters used in our analysis are also set according to the technical design 
report for ILC ~\cite{Behnke:2013xla} and we tabulate them in Table \ref{table:accelparam} for convenience.

\begin{figure*}[tbh]
        \centering
\includegraphics[width=7.2cm,height=6.5cm]{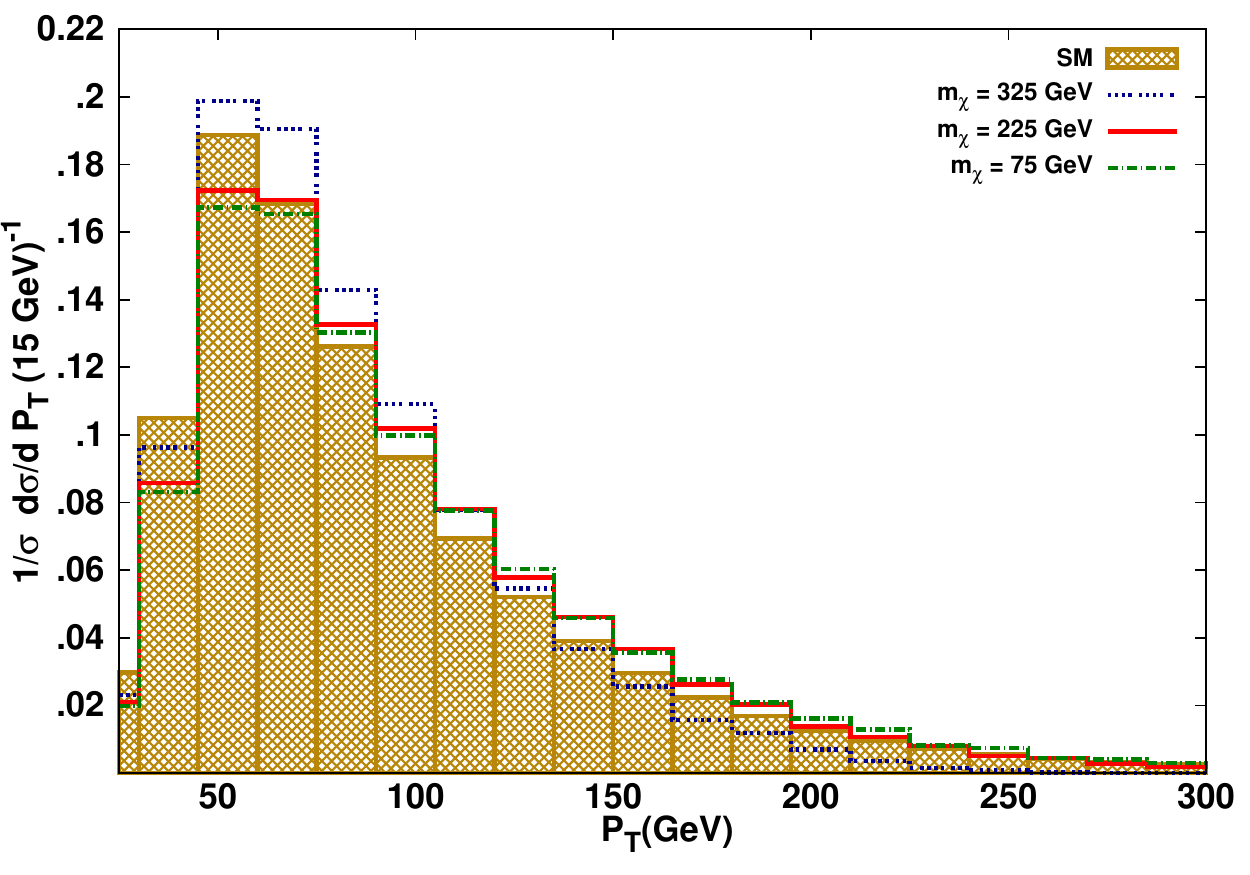}    
\includegraphics[width=7.2cm,height=6.5cm]{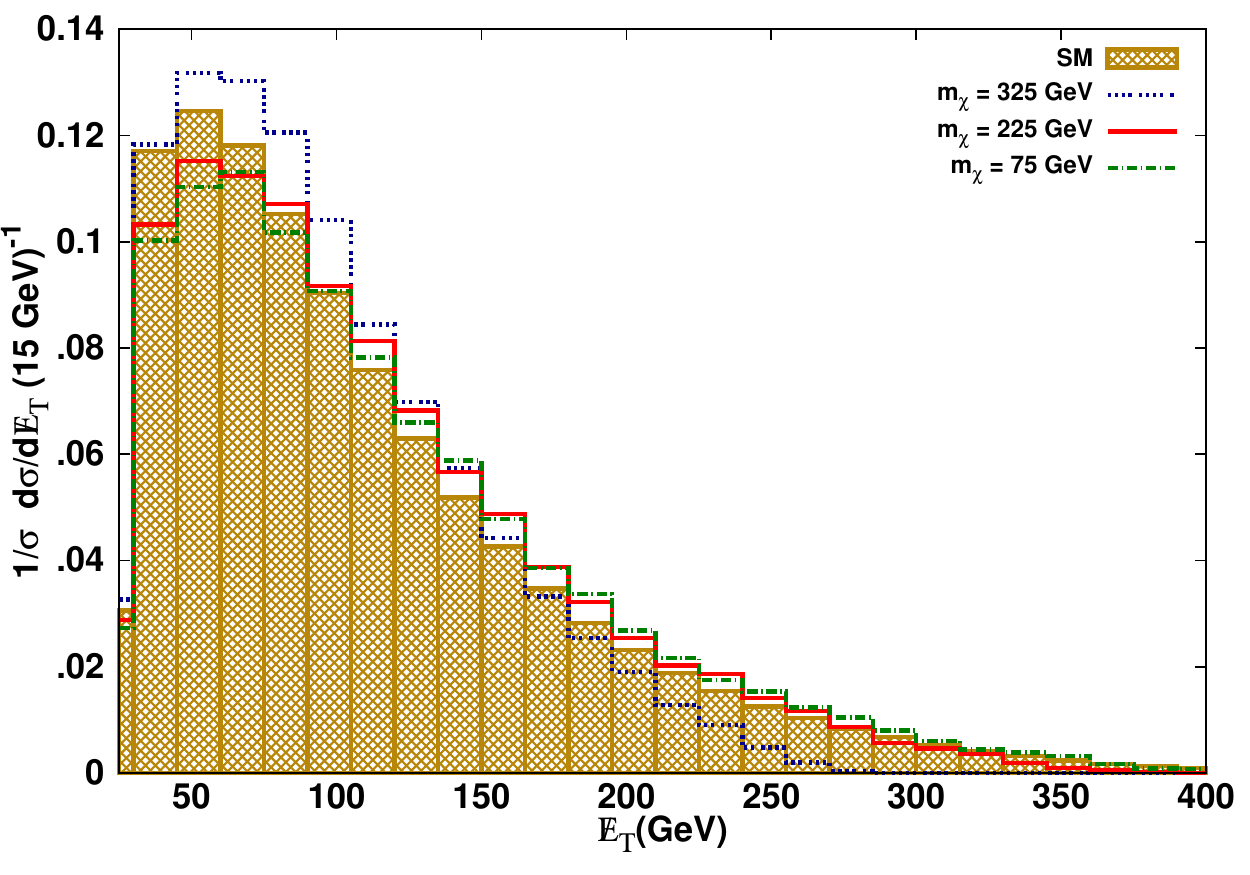}
\includegraphics[width=7.2cm,height=6.5cm]{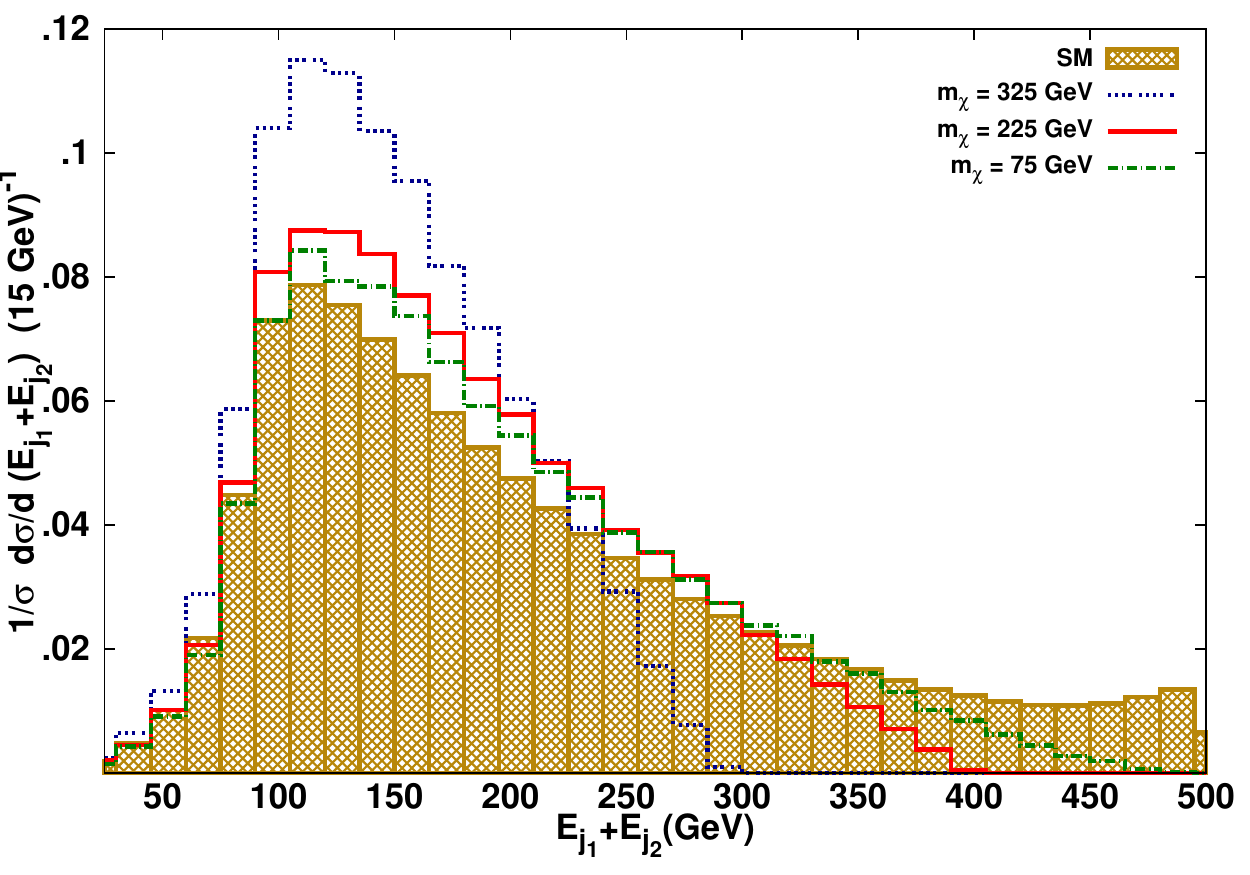}        
\includegraphics[width=7.2cm,height=6.5cm]{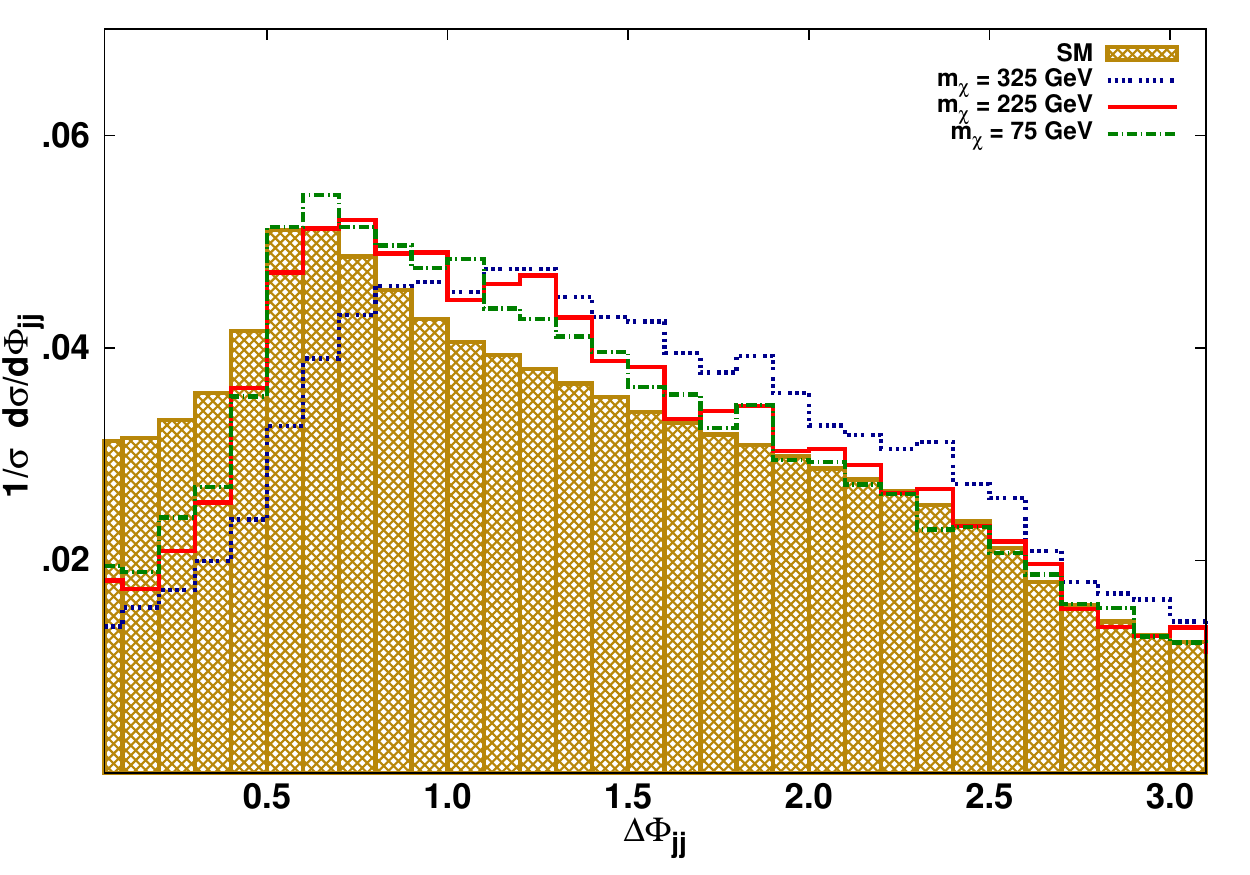}
        \caption{\small \em{Normalized 1-D differential distribution
        of the cross-section {\it w.r.t.} kinematic observables
        ${P_T}_{j_1}$, $\not \!\! E_T$, sum of energy of jet pairs and
        $\Delta\phi_{jj}$ for the background process (shaded
        histograms) and the $Z\to j\, j$ associated DM pair production
        at the three representative values of $M_{\chi}$ = 75 and
        325 GeV at $\sqrt{s}$ = 1 TeV, $\Lambda$ =1 TeV and $g_{VV}^l$
        ($g_{AA}^l$, $g_{VA}^l$, $g_{AV}^l$) = 1. The bin width is 0.1
        for the $\Delta\phi_{ii}$ distribution while it is 15 GeV for
        the $\not \!\! E_T$ and $E_{j_1}+E_{j_2}$. }}
        \label{fig:distrVpgs}
\end{figure*}

Before we commence an analysis of the cross sections and the
kinematic distributions we must impose  basic (acceptance) cuts on 
transverse momenta ($p_{T j/\ell}$), rapidities ($\eta_{j/\ell}$) 
and separation ($\Delta R_{jj/\ell\ell}$) of the visible entities 
(jets and  $\mu^\pm$, $e^\pm$) along with the the missing transverse
energy ($\not \!\!\! E_T$), which commensurate with the typical detector
capabilities on the one hand, and theoretical considerations 
(such as jet definitions) on the other. 
\begin{itemize}
 \item  $p_{T_{i}} \geq$10 GeV where $i=\ell,j$ ,  

 \item  $\left\vert\eta_\ell\right\vert\leq$ 2.5 and $\left\vert\eta_j\right\vert \leq$ 5,
 \item  $\Delta R_{ii} \geq 0.4$ where $i=\ell,j$ \ ,
 \item  $\left\vert M_{ii}-m_Z\right\vert\le 5\,\Gamma_Z$ where $i=\ell,j$; for the hadronic channel, this invariant mass actually refers to that for all the jets together,
 \item  $\not \!\! E_T \geq$25 GeV.
\end{itemize}
The corresponding cross-sections with these basic cuts for the polarized 
and unpolarized beams of electron and positron  are summarized in Table
~\ref{table:cs}. Feynman diagrams consisting of different topologies and
contributing to the background~\eqref{eqn:bkgd1} can be further suppressed 
by imposing the most-dominant  invariant-mass cut of the visible particles
\footnote{For $e^+ e^- \rightarrow e^+ e^- + \sum_{i} \nu_i\bar{\nu_i}$, 
the deviation from such a dominance is quite significant. Precisely for 
this reason, this turns out to be to be the least sensitive of the three 
channels.}, and thus allowing $Z +Z^{(*)}$ final state with the on-shell
$Z$ decaying into the two visible entities, and the off-shell $Z^{(*)}$
going to a neutrino-pair. Within this approximation, the diagrams 
contributing to the SM background processes~\eqref{eqn:bkgd1}-\eqref{eqn:bkgd3}
have identical topologies and differ only in the coupling of the $Z$ 
to the final state fermions. Consequently, this difference would
manifest itself essentially only in the total rates with differences
in the normalized distributions being of a subleading nature. This is 
true even for the signal events corresponding to these final states.
Hence we have displayed the one-dimensional normalized differential 
cross-section for  process with only jets as final state in Figures
~\ref{fig:distrSpgs} and~\ref{fig:distrVpgs}.

\begin{figure*}
\centering      
\includegraphics[width=8.5cm,height=6cm]{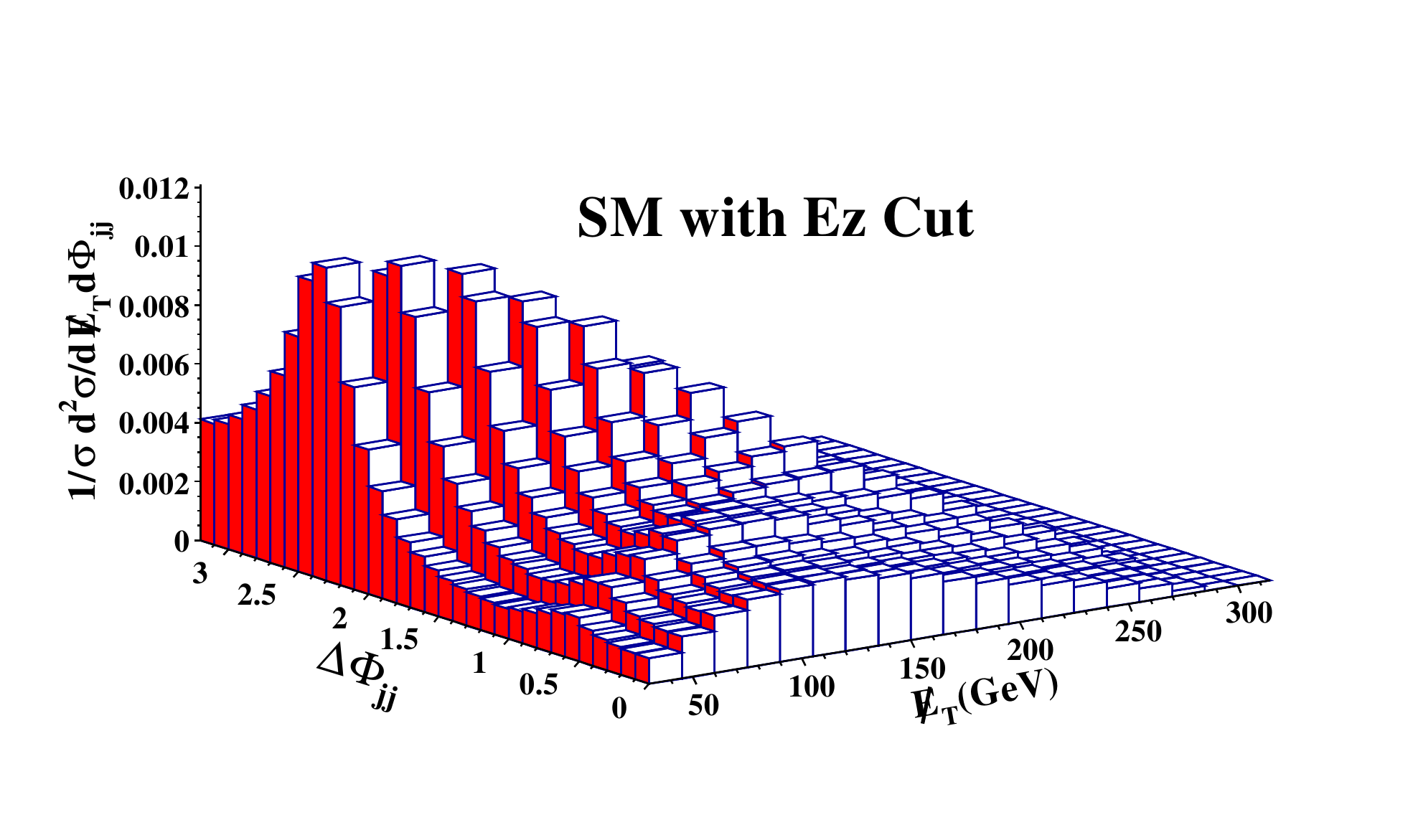}
\includegraphics[width=8.7cm,height=6cm]{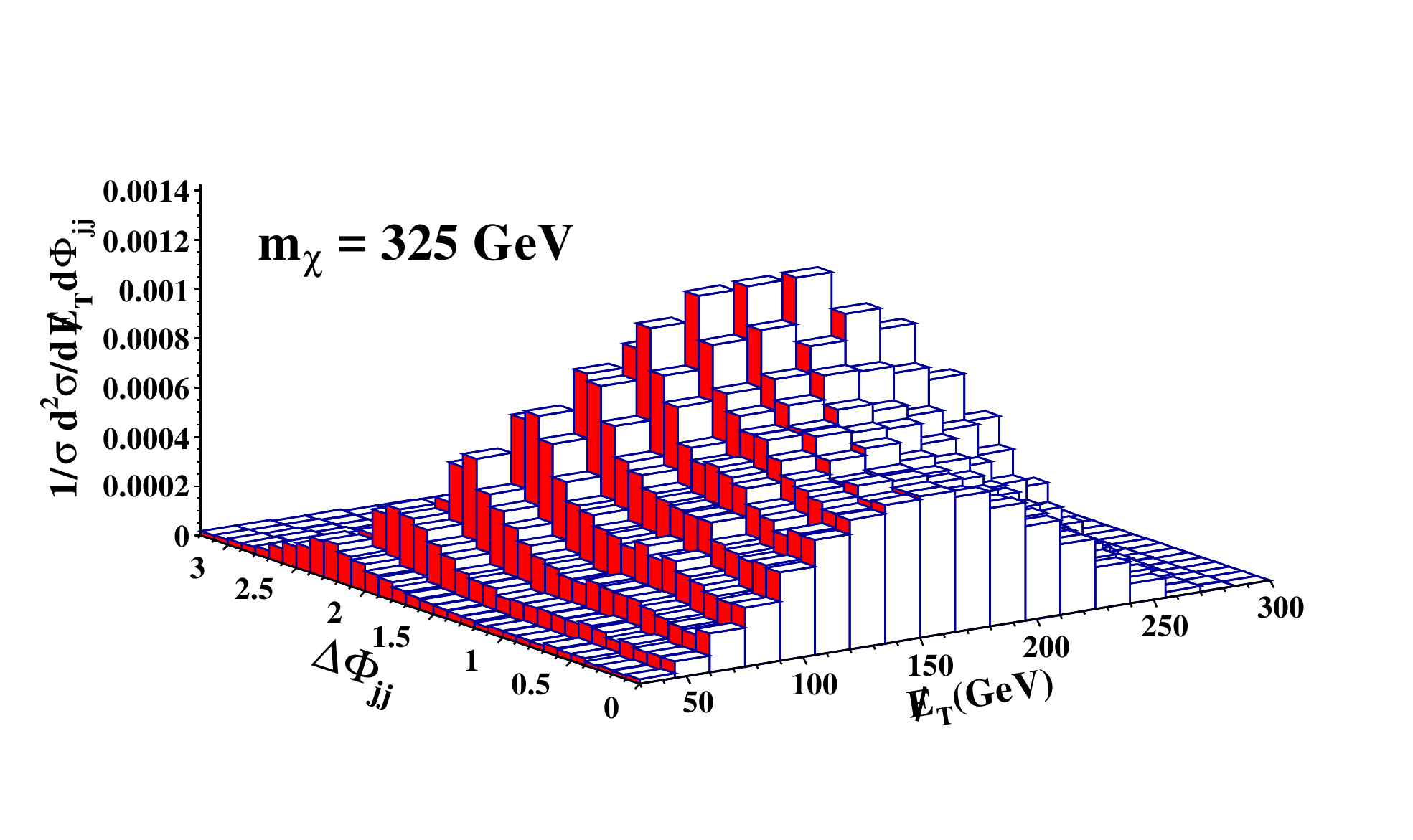}                   
        \caption{\small \em{ Lego plots showing normalized 2-D
        differential cross-section distribution {\it w.r.t.} $\not \!\! E_T$ and
        $\Delta\phi_{jj}$  for the background and
        the $Z\to j \,j$ associated DM pair production for $M_{\chi}$ =325 GeV at
        $\sqrt{s}$ = 1 TeV, $\Lambda$ =1 TeV and $g_{SS}^l$ = 1. The
        bin width is 15 GeV for the $\not \!\! E_T$ and 0.1 for the $\Delta\phi_{ii}$.}}
        \label{fig:2d}
\end{figure*}
\subsection{Analysis with one and two dimensional distributions:-}
\par In order to augment the signal-to-noise ratio by imposing additional 
selection cuts on the kinematic observables, we recourse ourselves to the
detailed study of the normalized one dimensional  differential cross-section
distributions {\it w.r.t.} for all the sensitive kinematic  observables 
{\it e.g.} $p_{T_{j/\,l}}$, $\not \!\!\! E_T$, $\Delta\phi_{ii}$ and sum of 
energies from a  jet or lepton pair $\left(E_{i_1}+E_{i_2}\right)$ for 
the signal with varying DM mass as well as varying effective DM pair 
- SM fermion pair couplings $g_{MN}^f/\Lambda^2$ at $\sqrt{s}$ = 500 GeV and 1 TeV. 
\par The very structure of the DM interaction Lagrangian
implies that, for a given final state, the normalized differential
distributions for the $SS$, $SP$, $PS$ and $PP$ cases are very similar
to each other, with the differences being proportional to the
difference in mass of the SM fermions in the final state (i.e., $e$,
$\mu$ or the light quarks). Similarly, the normalized distributions
for the $VV$, $VA$, $AV$ and the $AA$ cases are, again, very similar
to each other. Therefore, we have chosen to display the
normalized histograms for the observables
$p_{T_{j/\,l}}$, $\not \!\!\! E_T$, $\Delta\phi_{ii}$ and 
$\left(E_{j_1}+E_{j_2}\right)$ from   signal processes at the three
representative values of the DM masses $m_\chi$ namely 75, 225 and 325
GeV in Figures \ref{fig:distrSpgs} and \ref{fig:distrVpgs} corresponding to
the scalar and vector interactions of the DM bilinears with SM fermion
pair at $\sqrt{s}$ = 1 TeV and $\Lambda$ = 1 TeV. We list our
observations as follows:-

\begin{figure*}[h!]
        \centering
                 \subfloat{Process 1 : $ e^- e^+ \rightarrow 2 $ jets + $\not \!\! E_T$} \\      
                 \includegraphics[width=7.2cm,height=6.5cm]{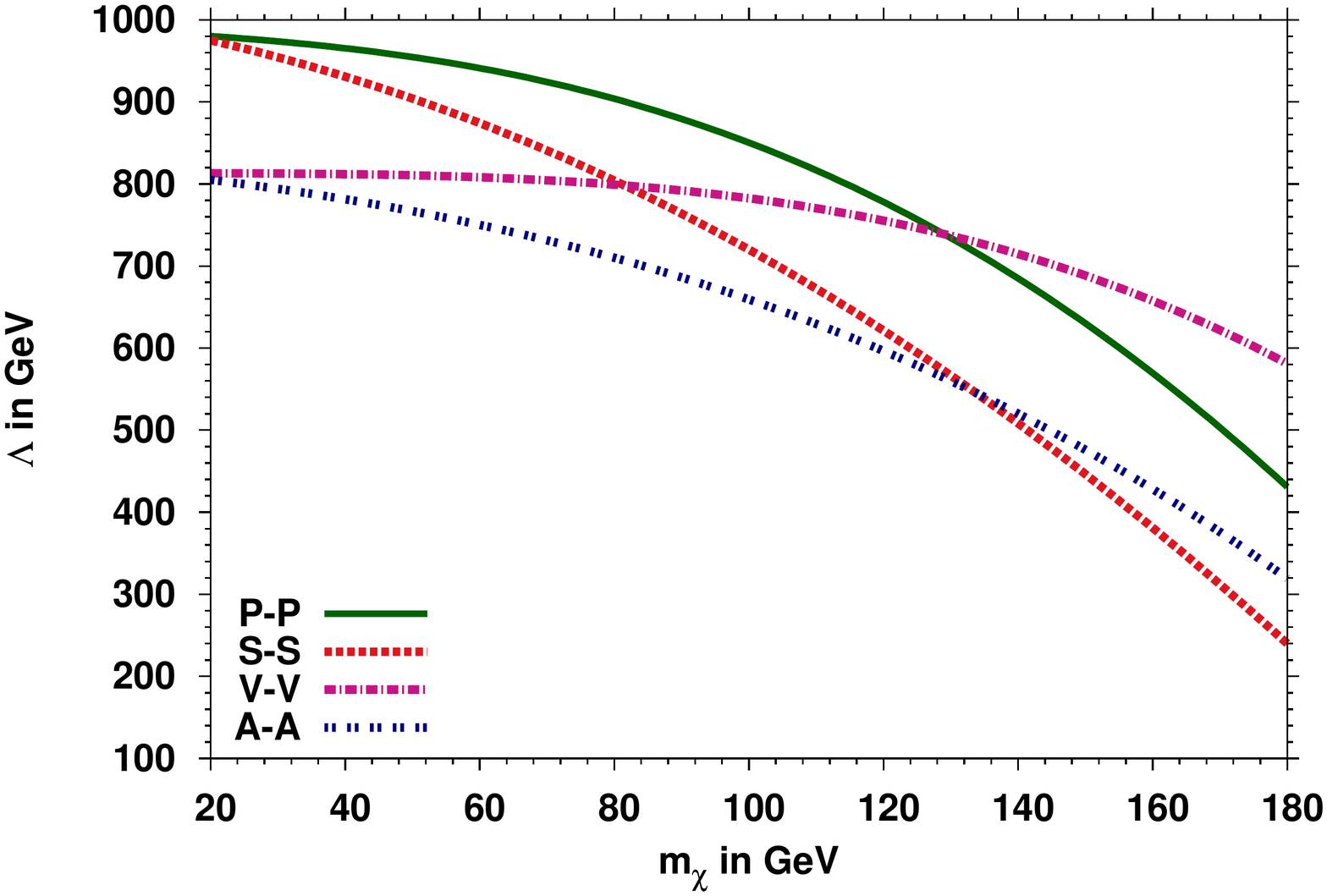}
                 \includegraphics[width=7.2cm,height=6.5cm]{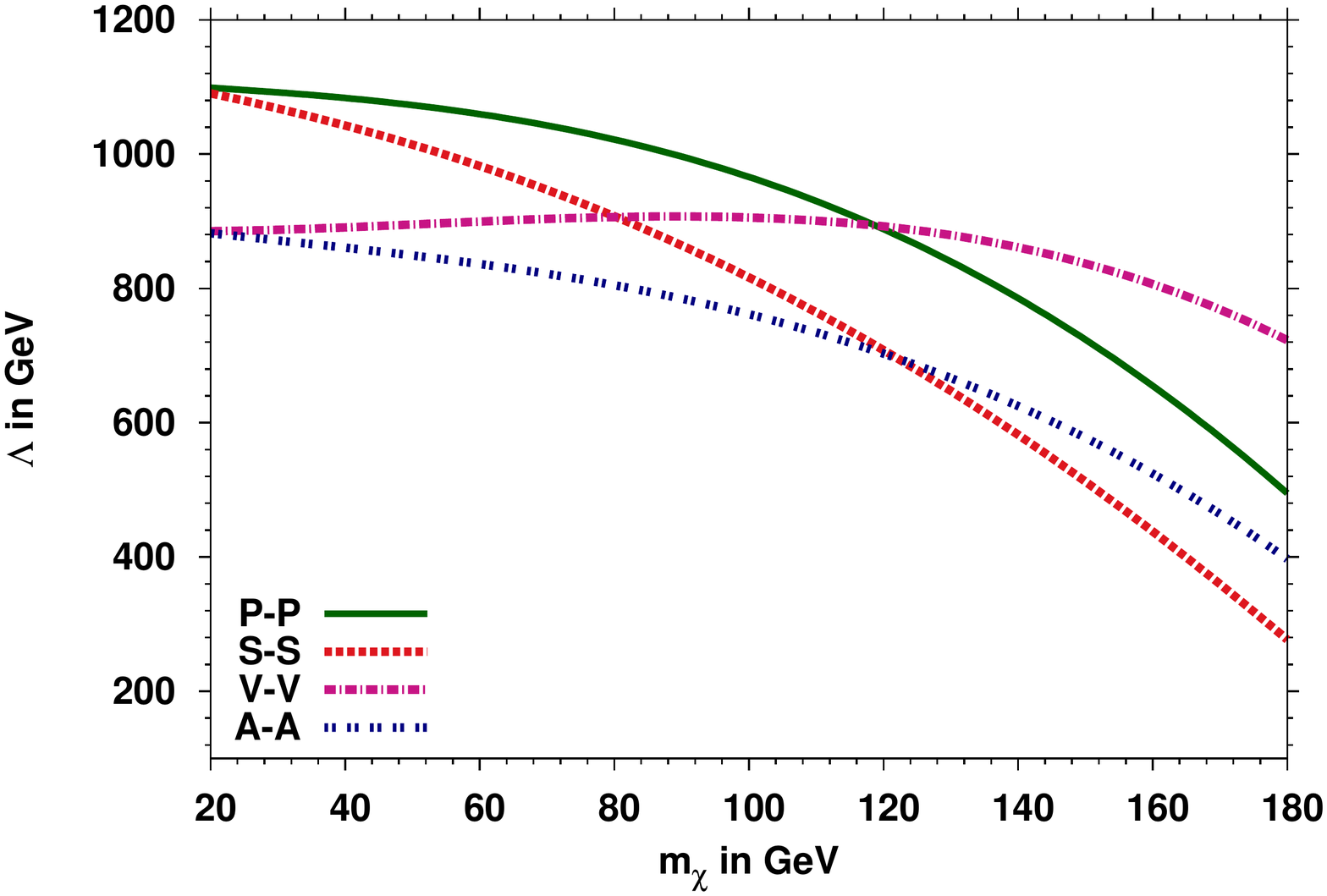}  \\      
                 \subfloat{Process 2 : $ e^+ e^- \rightarrow \mu^+ \mu^-$  + $\not \!\! E_T$}\\
                 \includegraphics[width=7.2cm,height=6.5cm]{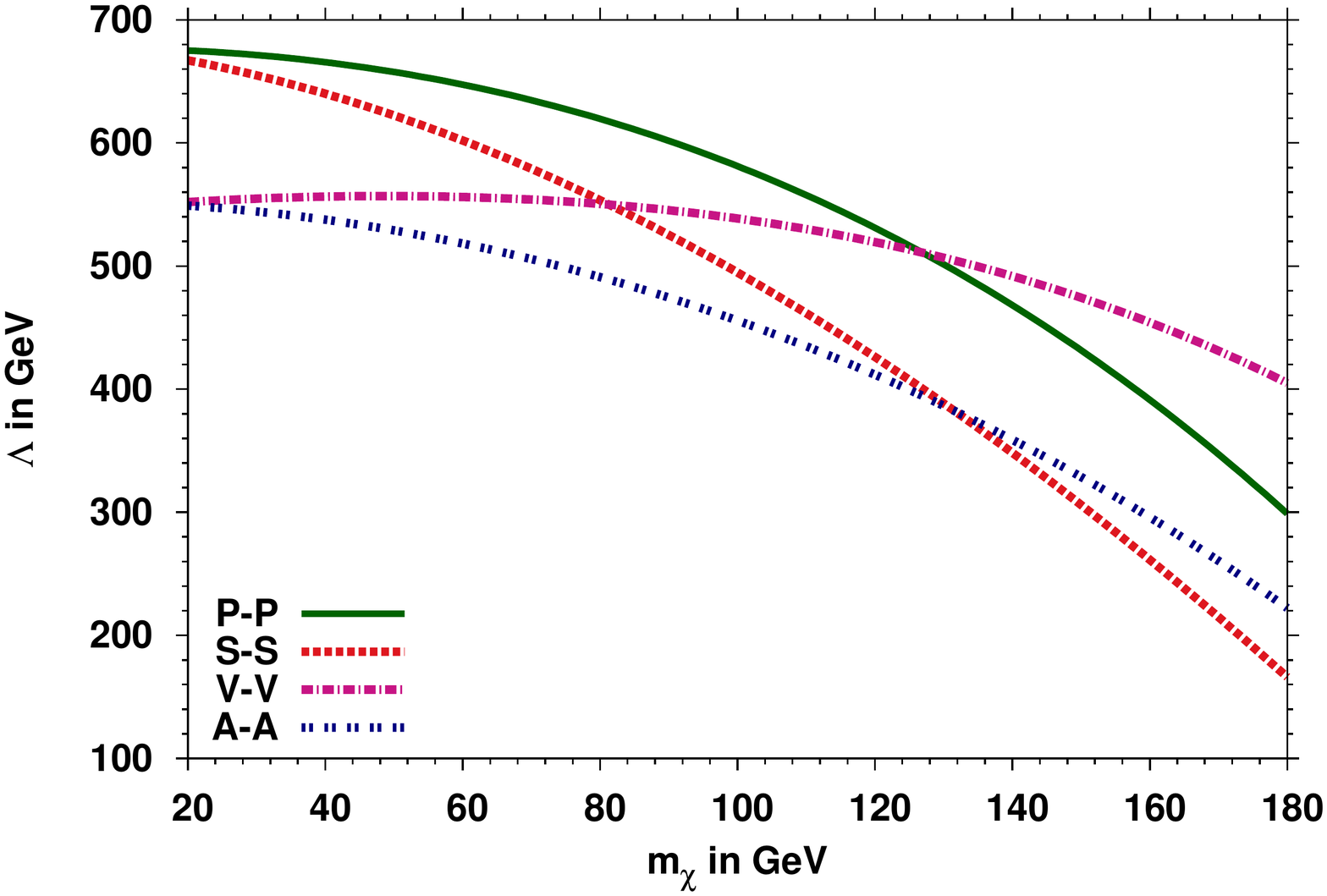}
                 \includegraphics[width=7.2cm,height=6.5cm]{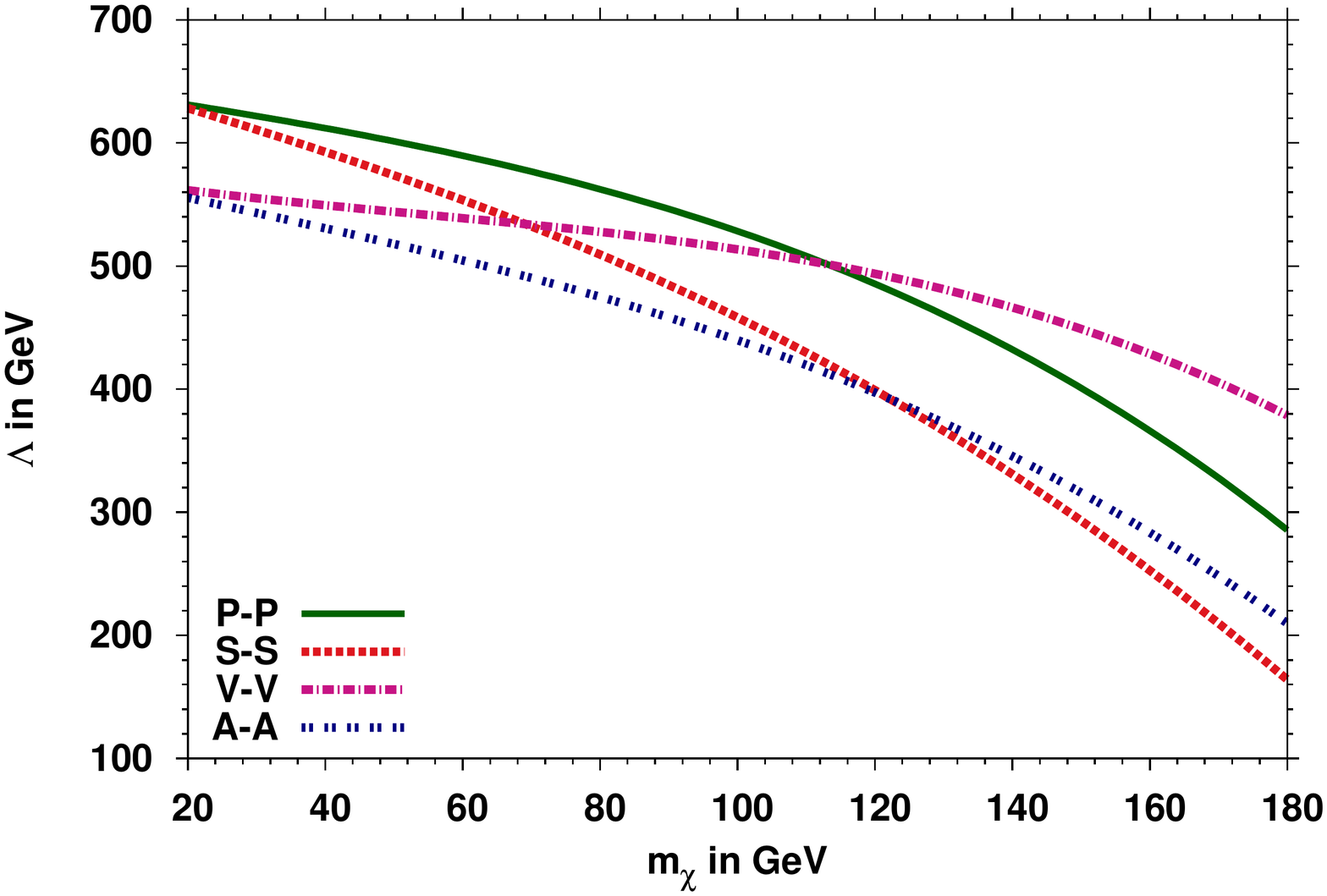}\\
                 \subfloat{Process 3 : $ e^+ e^- \rightarrow e^+ e^-$  + $\not \!\! E_T$}\\
                 \includegraphics[width=7.2cm,height=6.5cm]{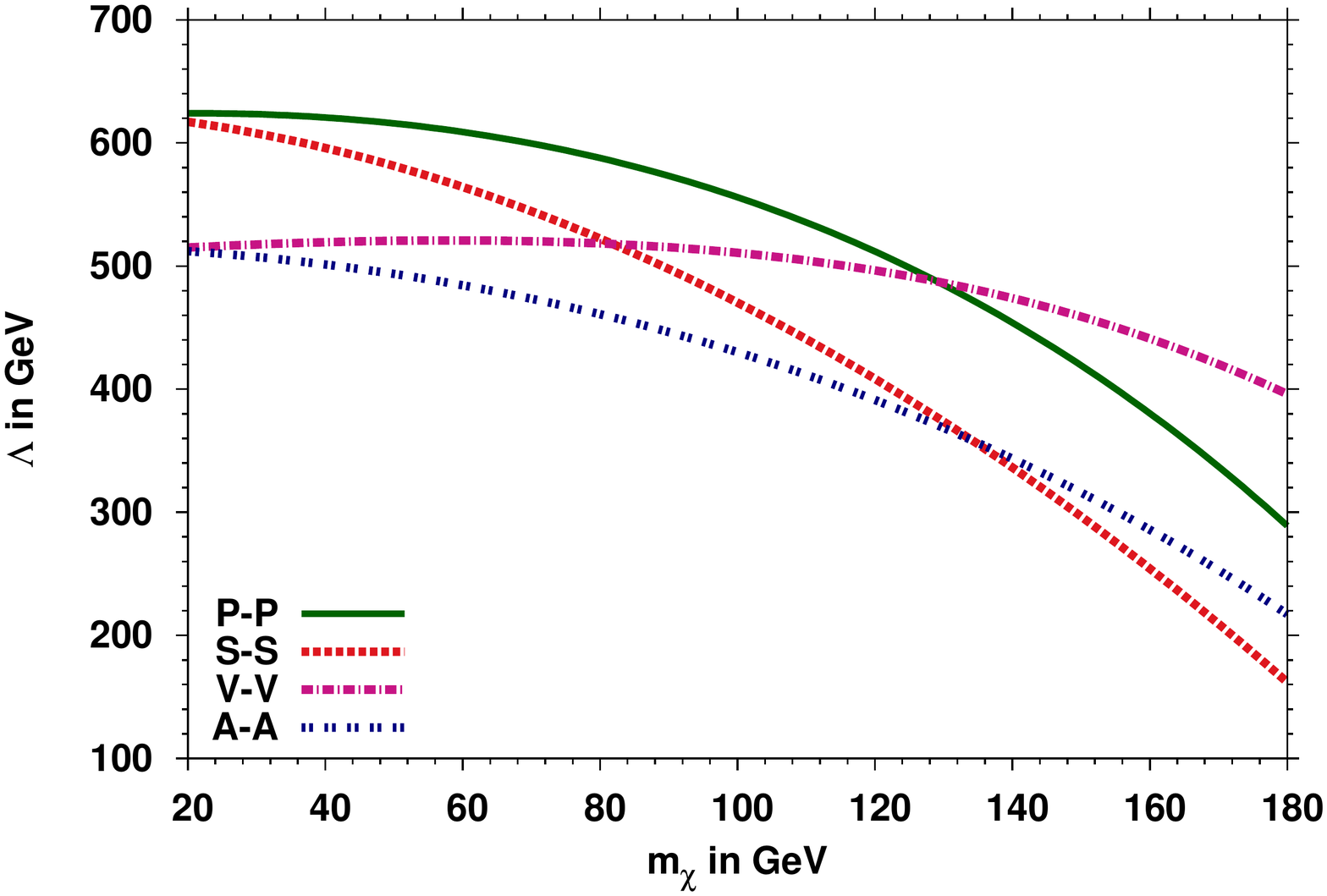}
                \includegraphics[width=7.2cm,height=6.5cm]{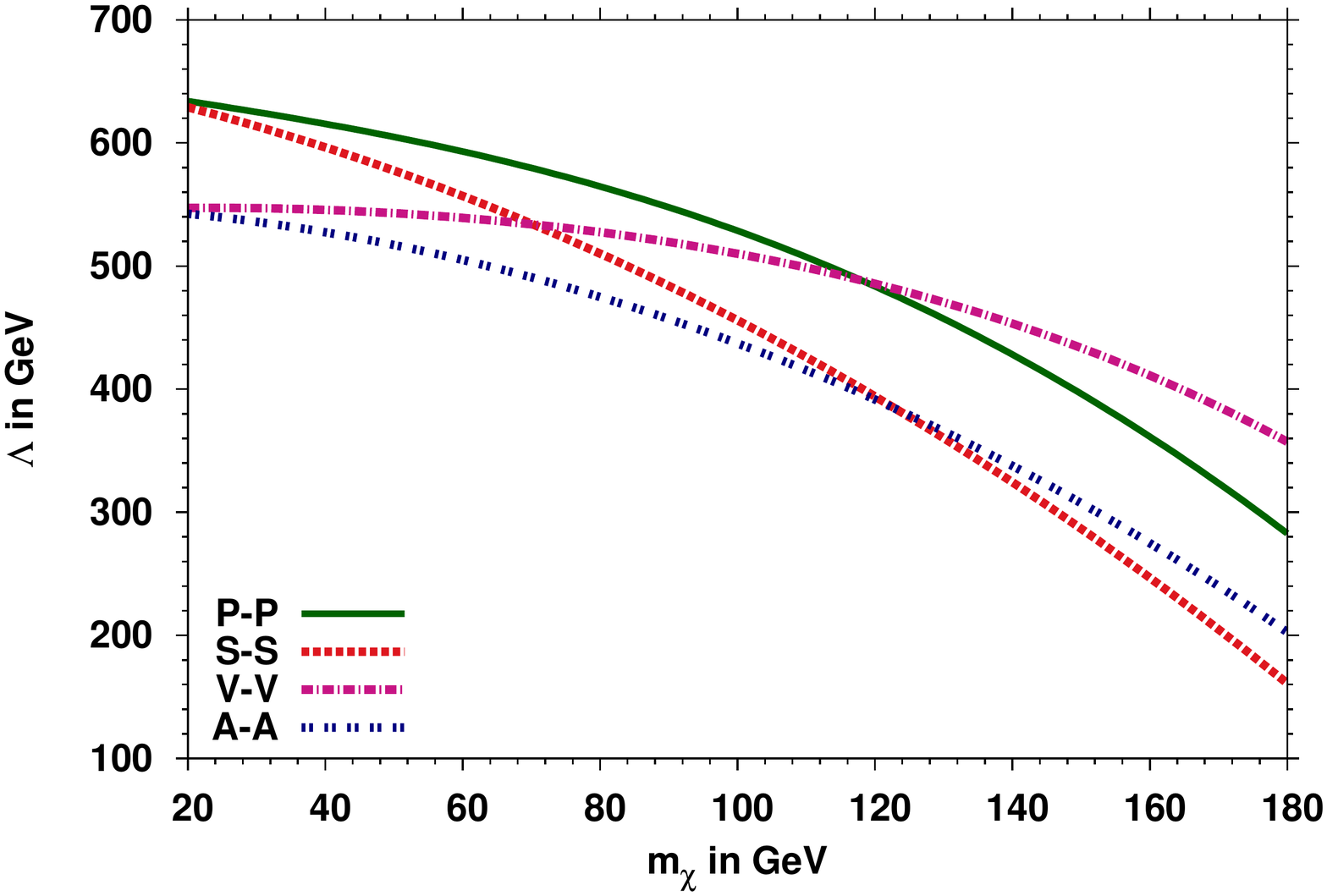}         
                  \caption{\small \em{ 99\% Confidence Level contours in the $m_\chi$ and $\Lambda$ plane  from the  $\chi^2$ analyses of the respective final  visible states at $\sqrt{s}$ = 500 GeV for the collider parameter choice given in  the first and the second column of the Table \ref{table:accelparam}. The  contours at the left  correspond to the unpolarized beam of $e^-$ and $e^+ $ with an integrated luminosity of 500 fb$^{-1}$. The contours at the right correspond to 80\%  and 30\% polarized beam for $e^-$ and $e^+$ respectively with an integrated luminosity of 250 fb$^{-1}$.}}
        \label{fig:contour_500_GeV}
\end{figure*}

\begin{figure*}[h!]
        \centering
                 \subfloat{Process 1 : $ e^- e^+ \rightarrow 2 $ jets + $\not \!\! E_T$} \\      
                 \includegraphics[width=7.2cm,height=6.5cm]{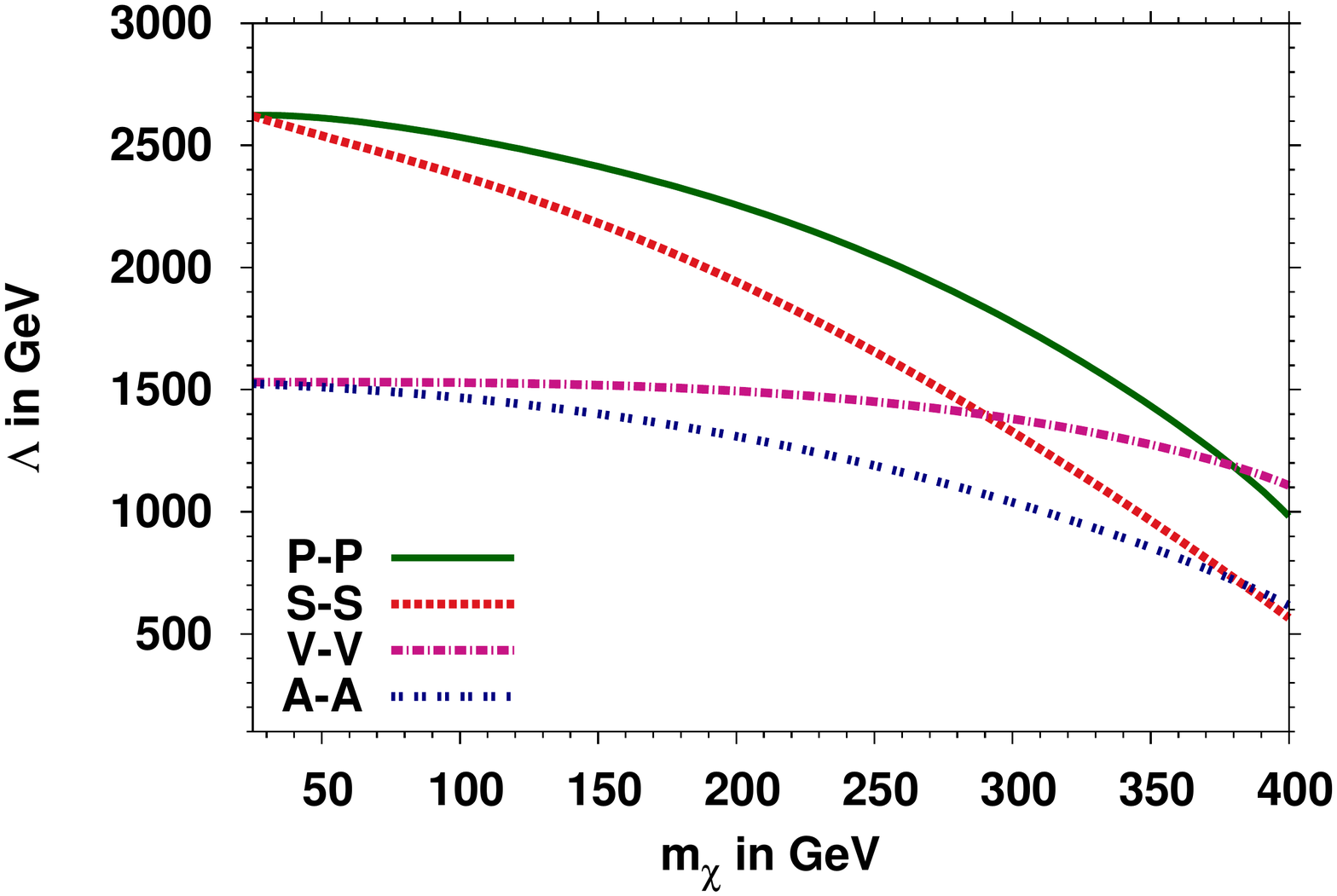}
                 \includegraphics[width=7.2cm,height=6.5cm]{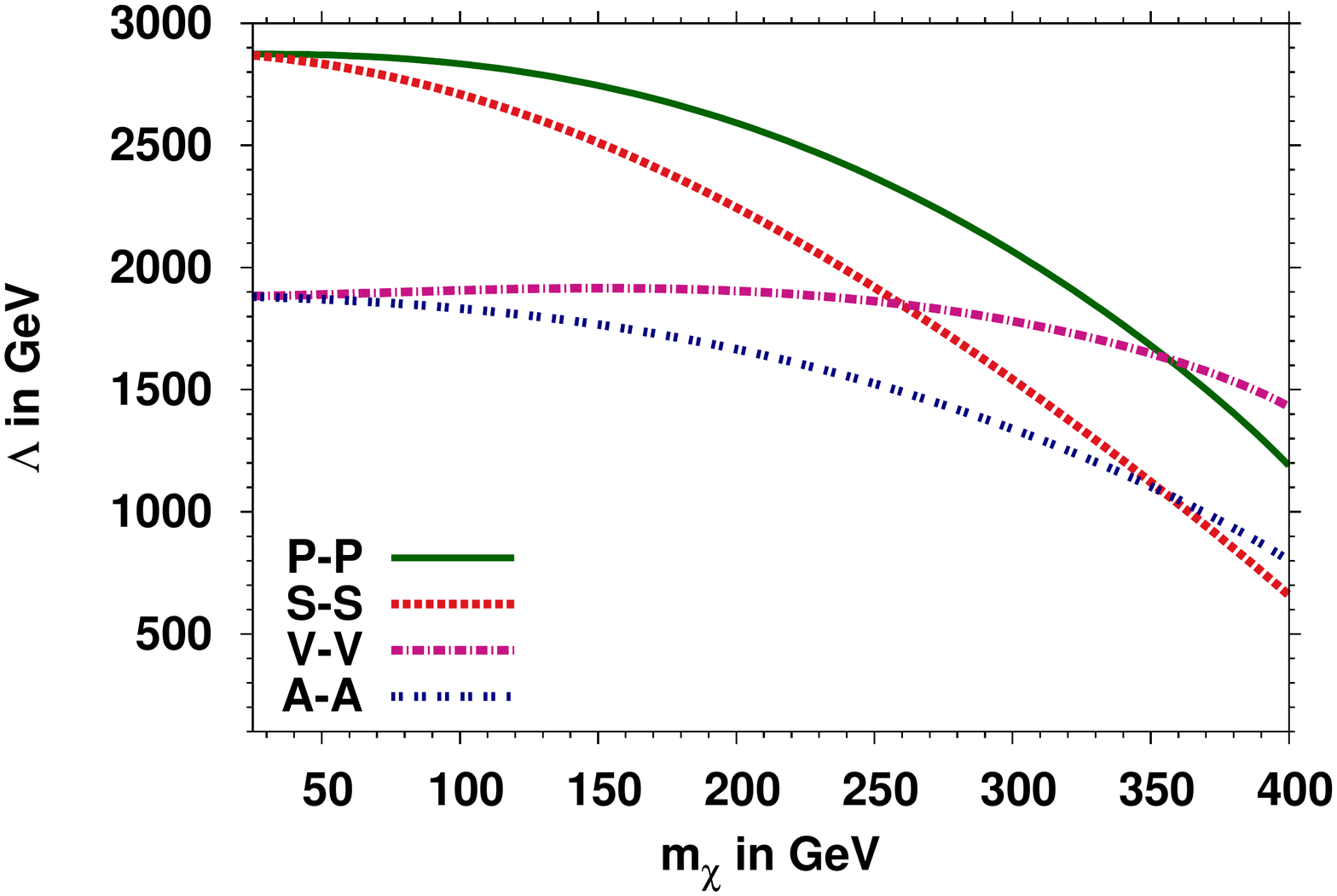}  \\      
                 \subfloat{Process 2 : $ e^+ e^- \rightarrow \mu^+ \mu^-$  + $\not \!\! E_T$}\\
                 \includegraphics[width=7.2cm,height=6.5cm]{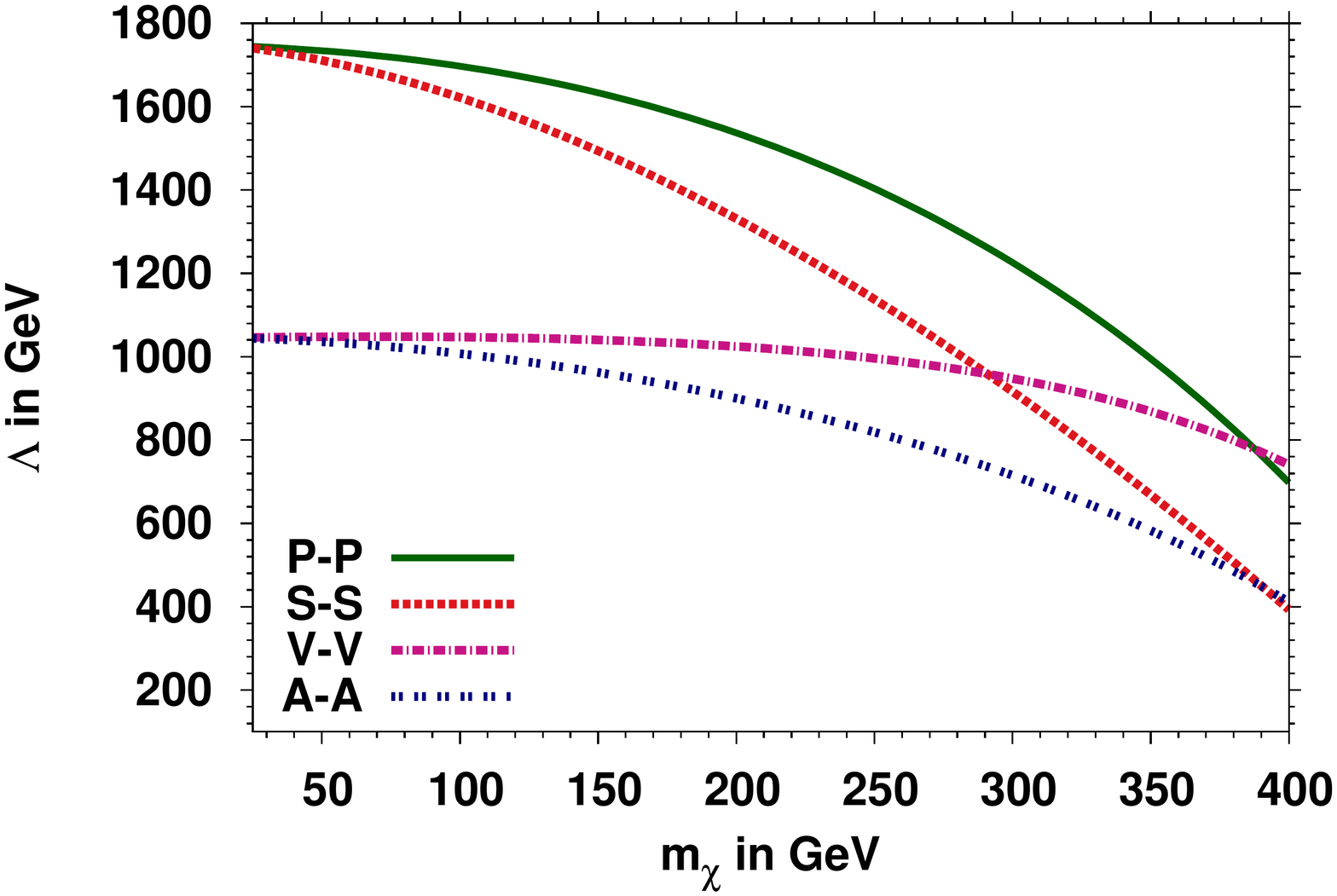}
                 \includegraphics[width=7.2cm,height=6.5cm]{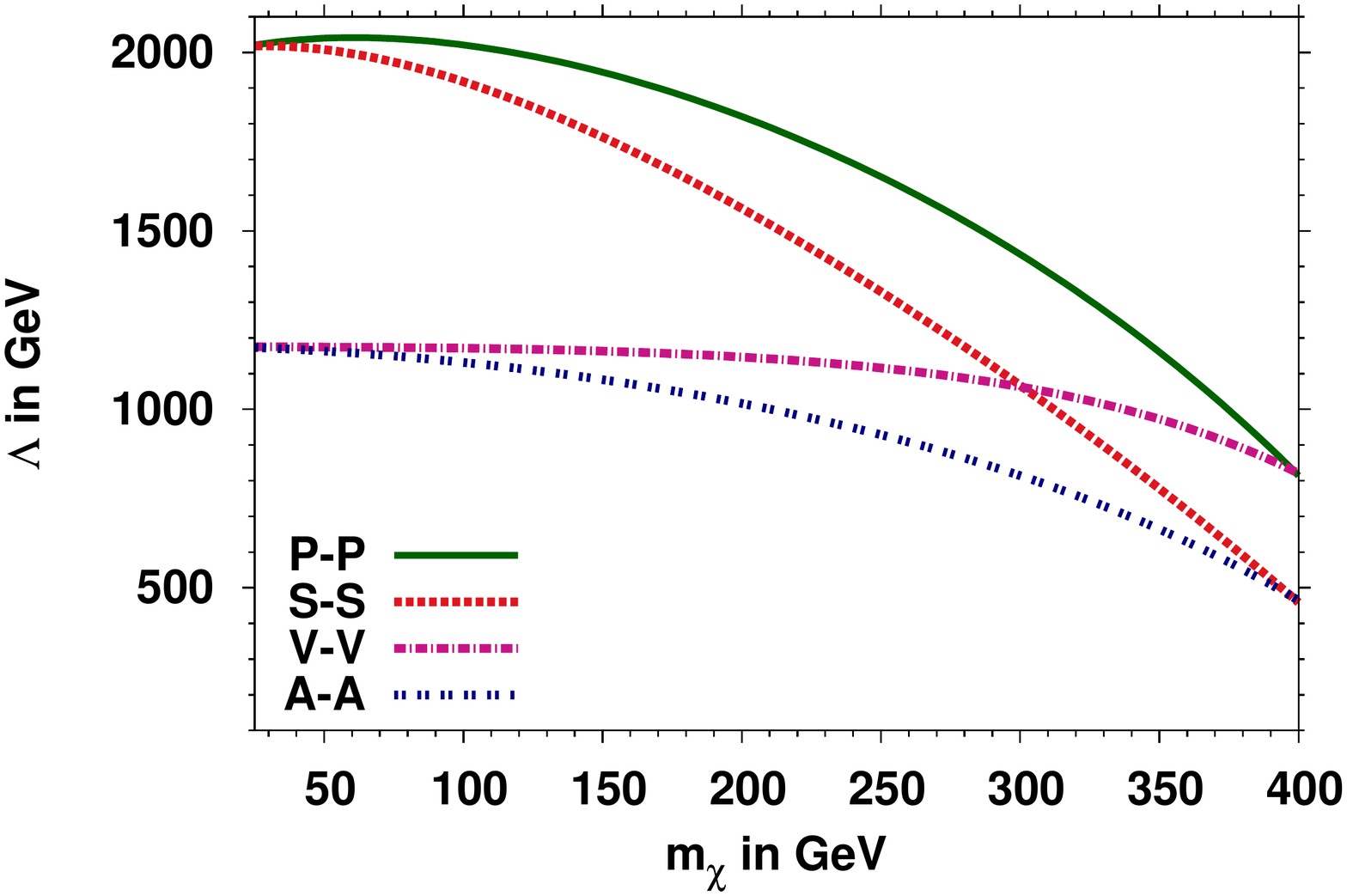}\\
                 \subfloat{Process 3 : $ e^+ e^- \rightarrow e^+ e^-$  + $\not \!\! E_T$}\\
                 \includegraphics[width=7.2cm,height=6.5cm]{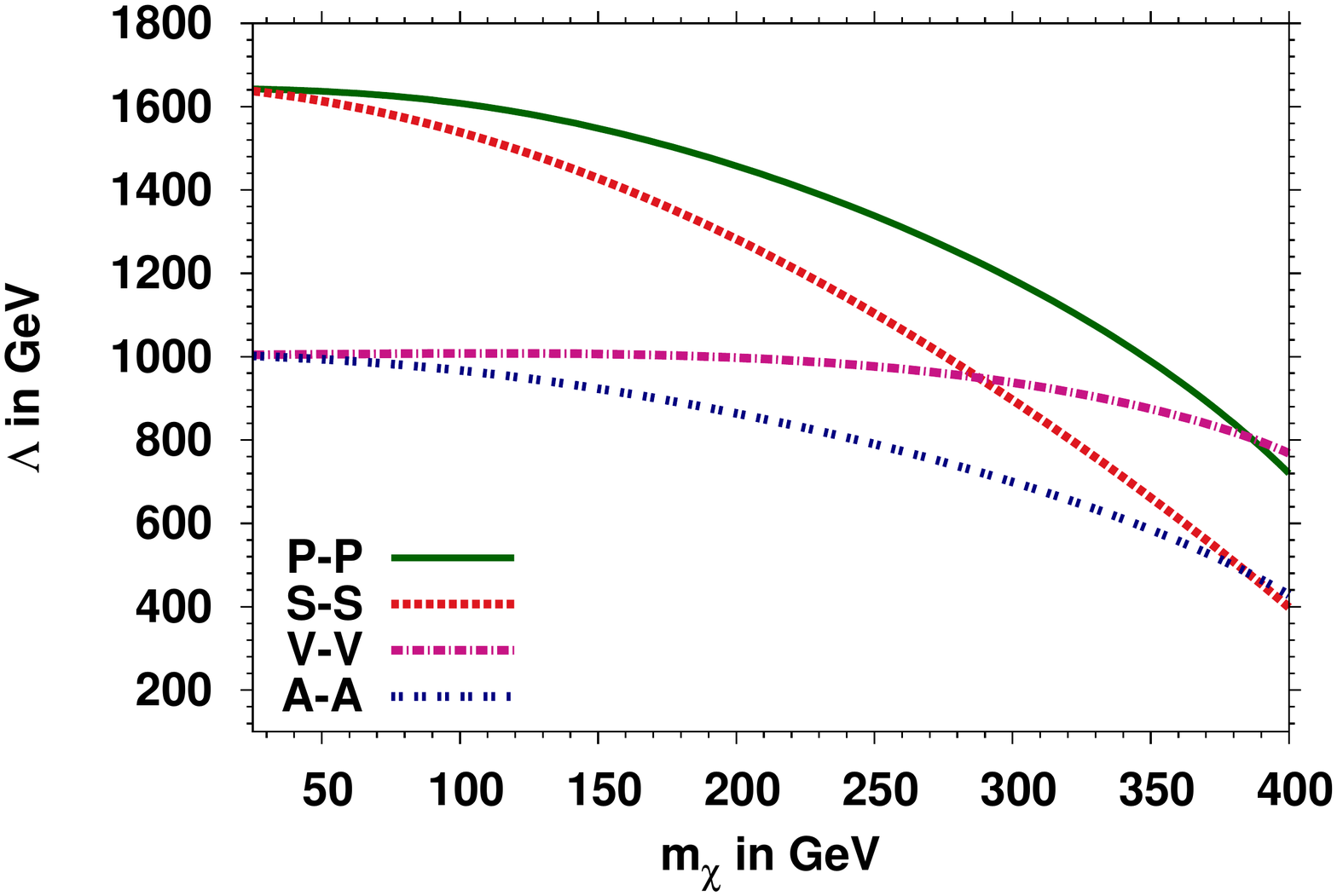}
                \includegraphics[width=7.2cm,height=6.5cm]{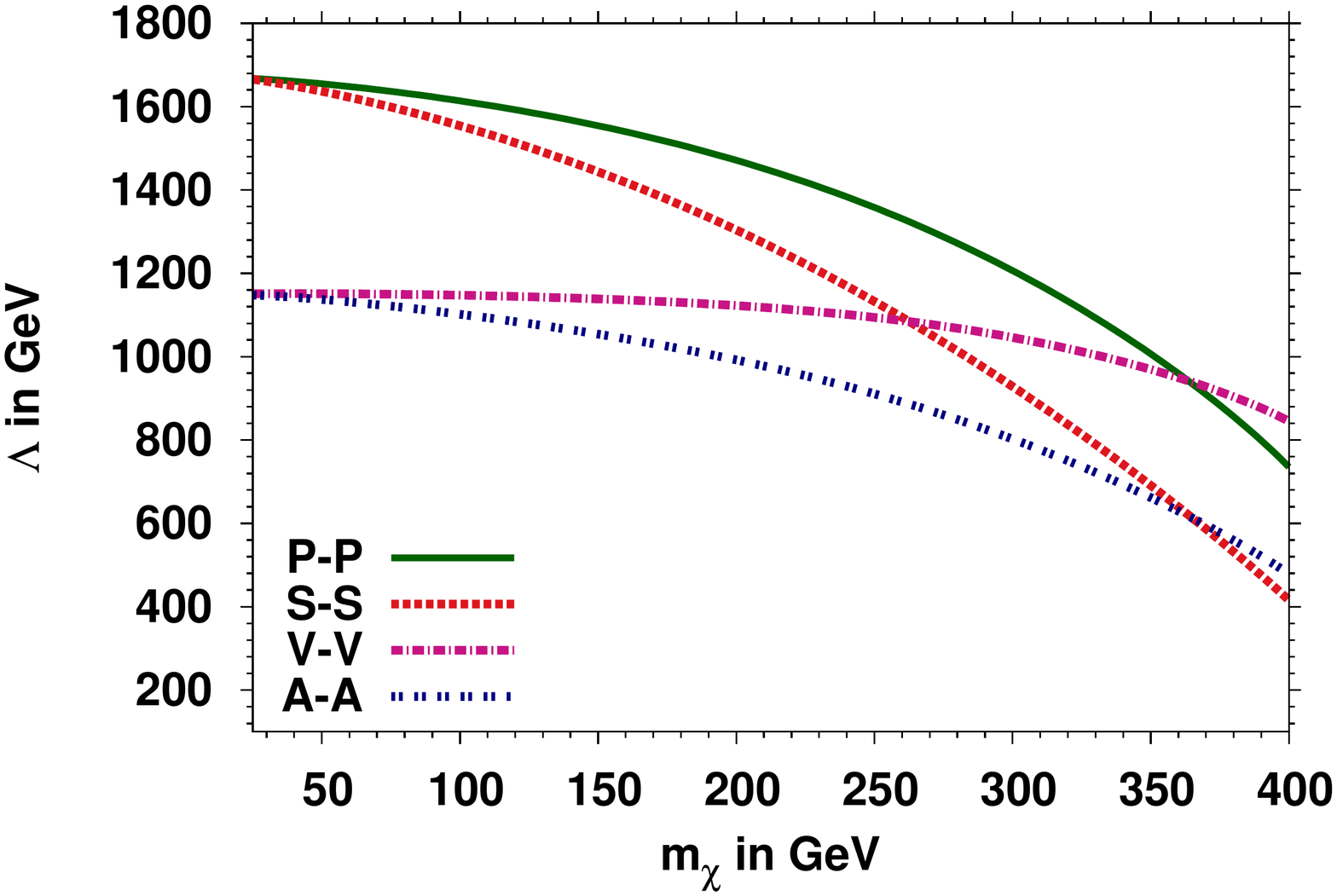}      
         \caption{\small \em{ 99\% Confidence Level contours in the $m_\chi$ and $\Lambda$ plane  from the  $\chi^2$ analyses of the respective final  visible states at $\sqrt{s}$ = 1 TeV for the collider parameter choice given in the  third and the fourth column of the Table \ref{table:accelparam}. The  contours at the left  correspond to the unpolarized beam of $e^-$ and $e^+ $ with an integrated luminosity of 1 ab$^{-1}$. The contours at the right correspond to 80\%  and 30\% polarized beam for $e^-$ and $e^+$ respectively with an integrated luminosity of 500 fb$^{-1}$.}}
        \label{fig:contour_1_TeV}
\end{figure*}
\begin{itemize}
\item[$\bullet$] In Figure \ref{fig:distrSpgs}, we observe that, as we increase the DM mass, the respective shape profile of  the all kinematic observables become more and more distinct {\it w.r.t.} the SM contribution. This difference in shape profile for the scalar coupling with that of SM can be attributed to the chirality-flip interactions of the DM which becomes  much more pronounced for higher values of $m_\chi$.
\par On contrary, the shape profiles of all the kinematic distributions with vector coupling of DM in Figure \ref{fig:distrVpgs}, behave the same as those of    SM  $Z\to\nu\bar\nu$ coupling. 

\item[$\bullet$] We find that in the lower $m_\chi$  region, the jet pair remains highly boosted as in the case of SM and hence, the peak of the differential cross-section distribution  {w.r.t.}  $\Delta\phi_{jj}$   coincides with that of SM background around $35^{\circ}$. The peak however shifts towards higher values of  $\Delta\phi_{jj}$ as we increase $m_{\chi}$. 

\item[$\bullet$] We find that the energy profile of the visible pair  $E_{j_1}+E_{j_2}$ are sensitive to the choice of the DM mass and thus advocating that the upper limit on this kinematic observable can be used as a dynamic mass dependent selection cut  to reduce the continuum  background. This is realized by invoking  a cut on the invariant mass of visible particles to be around $\left\vert \left(p_{j_1}+p_{j_2}\right)^2- m_Z^2\right\vert^{\frac{1}{2}}\le 5 \,\Gamma_Z$, where $\Gamma_Z=2.49$ GeV. This restriction  translates  to an upper limit on the variable $E_{j_1}+E_{j_2}$ as a function of $m_\chi$: 
\begin{equation}
 E_{j_1}+E_{j_2} \leq \frac{s+m_Z^2-4m_{\chi}^2}{2\sqrt{s}}.
\label{eqn:EZcut}
\end{equation}
\end{itemize}

\par Since we are primarily interested in $\bar\chi \chi$ production associated with on-shell $Z\to jj\,/ ll$, we shall use the condition on the total visible energy ~\eqref{eqn:EZcut} as the Selection cut for the rest  of   our analysis. The signatures where a pair of leptons appear in the final state  due to the alternative decay modes of  $Z$ boson, the differential cross-section distributions {w.r.t.} the observables  $\Delta\phi_{l^+l^-}$ and $E_{l^+}+E_{l^-}$ are similar to the observables  $\Delta\phi_{jj}$ and $E_{j_1}+E_{j_2}$ which are  obtained from the respective  $\not \!\! E_T \,\, + 2\, {\rm jets}$. Therefore, we adopt the similar selection cut criteria on the $\not \!\! E_T$ and $E_{l^+}+E_{l^-}$.
\par Based on the study of 1-D differential cross-section distributions at the given luminosity we select the two most sensitive kinematic observables say $\not \!\! E_T$ and either $\Delta\phi_{jj}$ or $ \Delta\phi_{l^+l^-}$  for the  further analysis of two dimensional differential cross-section distributions. We analyse the efficiency  of the Selection cut  ~\eqref{eqn:EZcut} through the double distributions of the  simulated $Z$ associated DM pair production  events  (for  $m_\chi$ values at 75, 225 and 325 GeV), which  are then compared  with the   double  differential cross-section distributions of the SM at the given integrated luminosity.

\par In Figure ~\ref{fig:2d}, we exhibit the normalized two dimensional  differential cross-section with the Selection cuts in the $\not \!\! E_T$ and $\Delta\phi_{jj}$ plane for  the background and  the DM with $m_\chi$ = 325 GeV. These 3D lego plots are   generated from MadAnalysis 5 \cite{Conte:2012fm} 
at $\sqrt{s}$ = 1 TeV, $\Lambda$ =1 TeV and $g_{SS}^q$ = 1. The bin width is 15 GeV for the $\not \!\! E_T$ while it is  0.1 for the $\Delta\phi_{ii}$ distribution. 
\par We observe that the suppression of background processes due to the implementation of selection cuts  enhances the sensitivity of the signal.  We study the signal  efficiency ($S  /\sqrt{S+B}$) at $\sqrt{s}$ = 1 TeV  with the integrated luminosity 1 ab$^{-1}$ and in  the Table ~\ref{table:figureofmerit} we compare the 3$\sigma$ reach of the cut-off $\Lambda$  at a given mass $m_\chi$ for the dominant signature $e^+e^-\to \not \!\! E_T+ 2\,{\rm jets}$ corresponding to  the three representative masses of DM.

\subsection{$\chi^2$ Analysis}
In this subsection we study the sensitivity of the cut-off scale $\Lambda$ with the DM mass keeping the couplings of the DM bilinears with the SM fermionic pair to be unity. The $\chi^2$ analysis is performed with the sum of the variance of the differential bin events at a given integrated luminosity due to the presence of the new physics (NP) contribution over the SM events.
\par We define the $\chi^2$ for double distribution as
\begin{eqnarray}
\chi^2 = \sum_{j=1}^{n_1}\sum_{i=1}^{n_2}\left ( \frac{N_{ij}^{NP}}{\sqrt{N_{ij}^{SM+NP}+\delta_{sys}^2(N_{ij}^{SM+NP})^2}} \right )^2
\end{eqnarray}
\begin{figure*}[h!]
        \centering 
         \includegraphics[width=7.2cm,height=6.5cm]{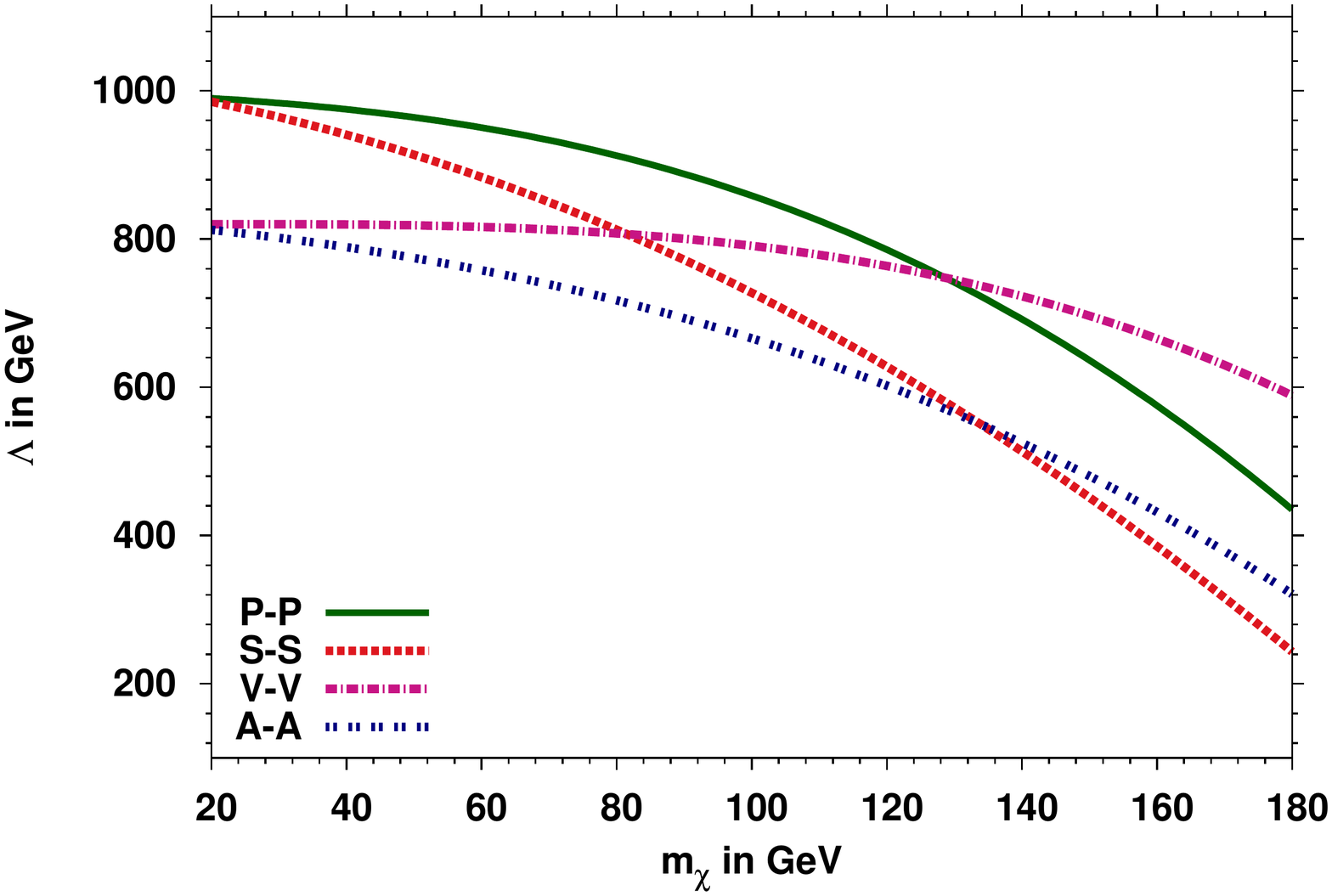}
         \includegraphics[width=7.2cm,height=6.5cm]{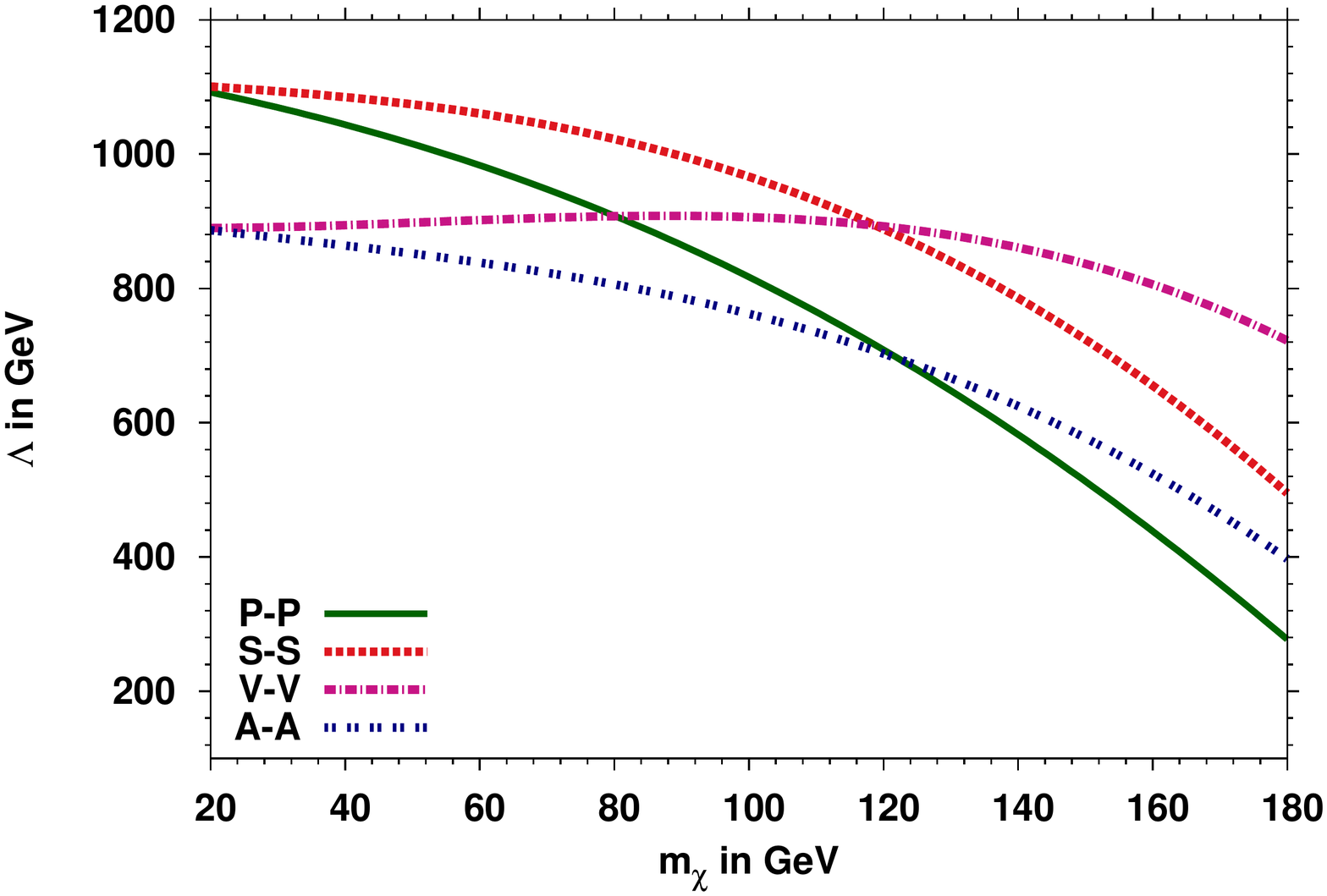}        
         \caption{\small \em{ 99\% Confidence Level contours in the $m_\chi$ and $\Lambda$ plane  from the combined $\chi^2$ analyses of all the three visible modes at $\sqrt{s}$ = 500 GeV for the collider parameter choice given in  the first and second column of the Table \ref{table:accelparam}. The  contours at the left  correspond to the unpolarized beam of $e^-$ and $e^+ $ with an integrated luminosity of 500 fb$^{-1}$. The contours at the right correspond to 80\%  and 30\% polarized beam for $e^-$ and $e^+$ respectively with an integrated luminosity of 250 fb$^{-1}$.}}
        \label{fig:contour_Comb_500_GeV}
\end{figure*}

\begin{figure*}[tbh]
        \centering        
                \includegraphics[width=7.2cm,height=6.5cm]{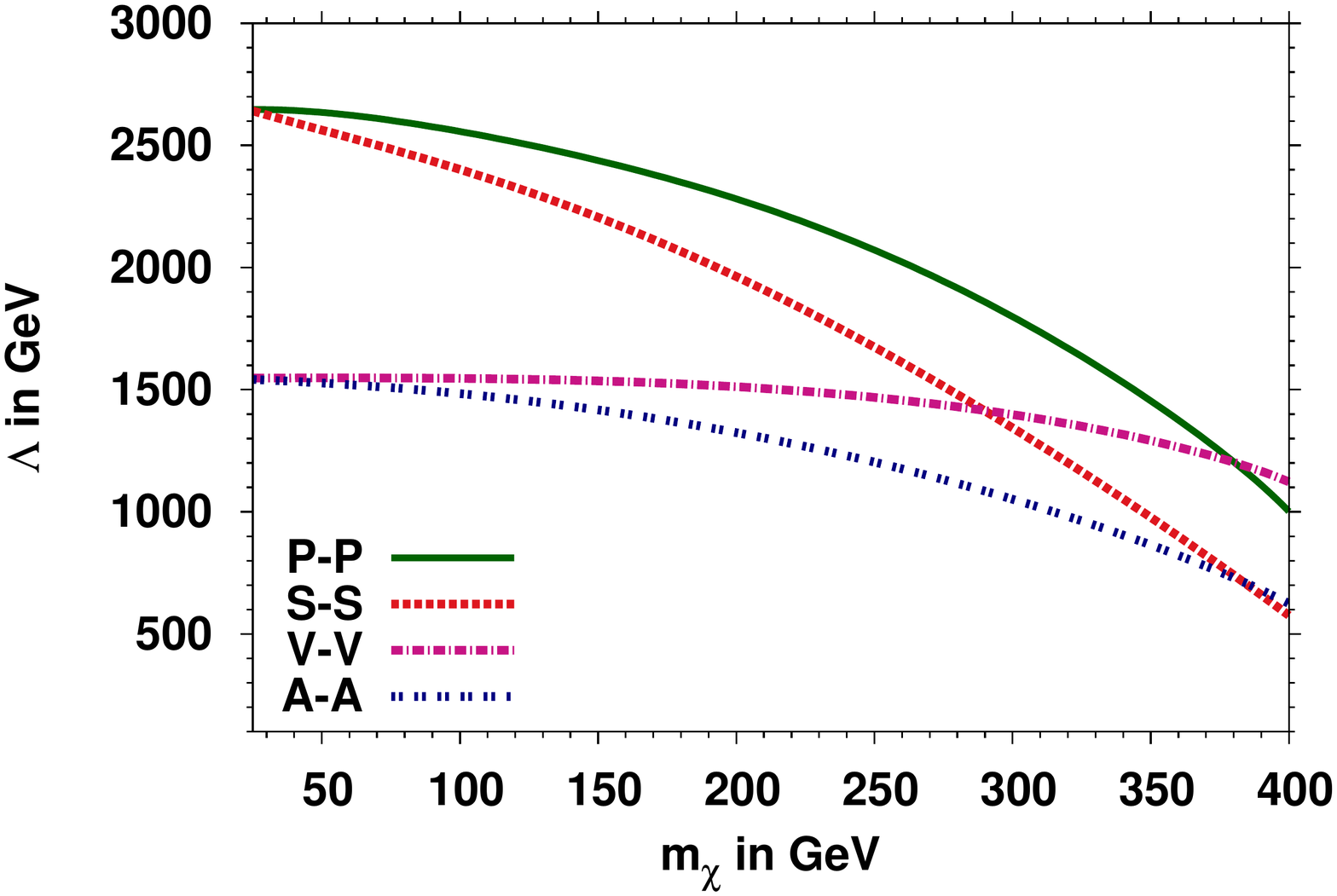}
                \includegraphics[width=7.2cm,height=6.5cm]{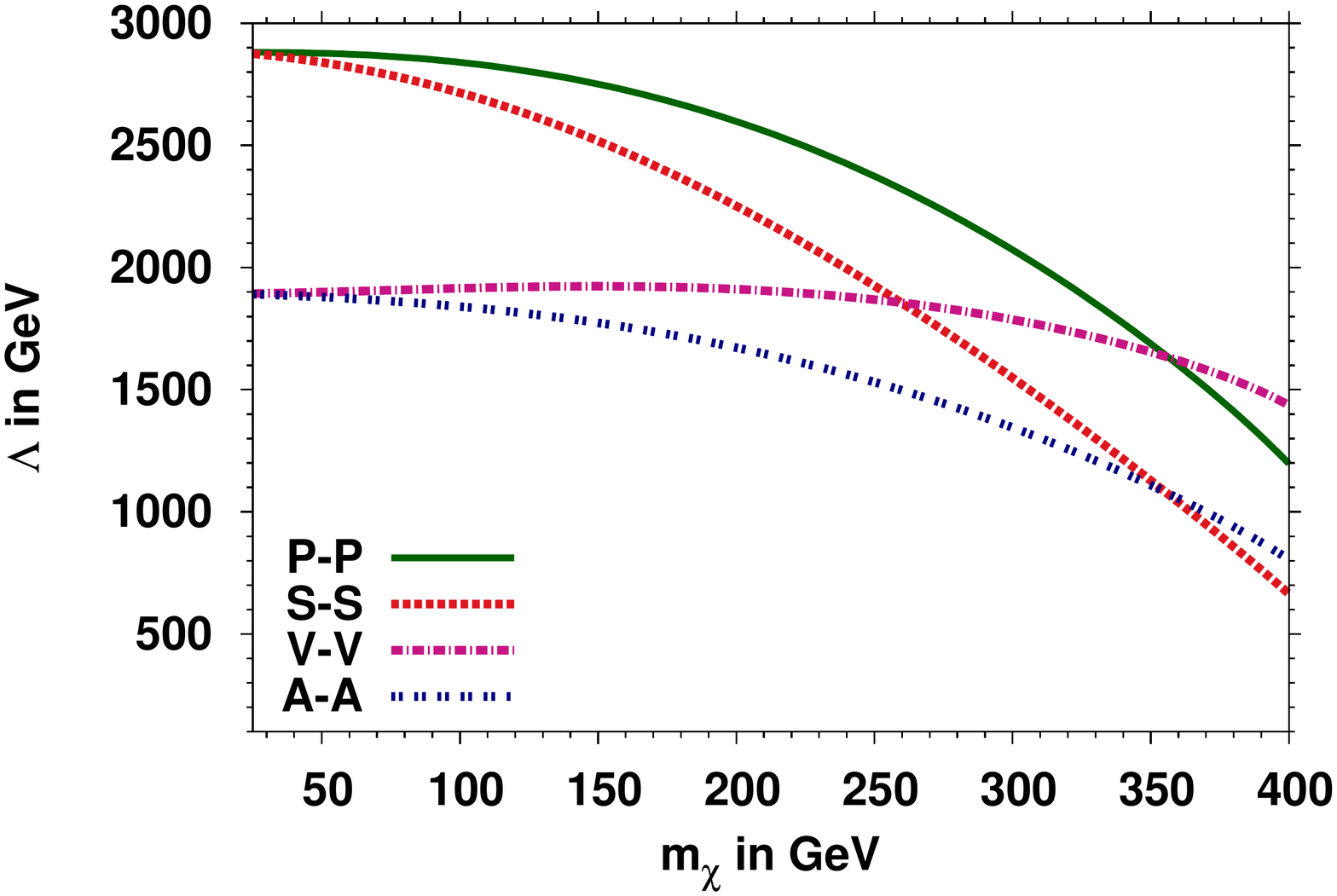}         
                 \caption{\small \em{ 99\% Confidence Level contours in the $m_\chi$ and $\Lambda$ plane  from the combined $\chi^2$ analyses of all the three visible modes at $\sqrt{s}$ = 1 TeV for the collider parameter choice given in  the third and fourth column of the Table \ref{table:accelparam}. The  contours at the left  correspond to the unpolarized beam of $e^-$ and $e^+ $ with an integrated luminosity of 1 ab$^{-1}$. The contours at the right correspond to 80\%  and 30\% polarized beam for $e^-$ and $e^+$ respectively with an integrated luminosity of 500 fb$^{-1}$.}}    \label{fig:contour_Comb_1_TeV}
\end{figure*}
\begin{table*}[ht]\footnotesize
\begin{center}
\begin{tabular}{c||cc|cc||cc|cc}
    \hline\hline
&\multicolumn{4}{c||}{\bf \underline{Unpolarized}}&\multicolumn{4}{c}{\bf \underline{Polarized}}\\

Interactions    &\multicolumn{2}{c|}{S-S}&\multicolumn{2}{c||}{V-V}&\multicolumn{2}{c|}{S-S}&\multicolumn{2}{c}{V-V}\\
 $\sqrt{s}$ in TeV   &0.5 &1&0.5&1 &0.5&1&0.5&1\\
${\cal L}$ in fb$^{-1}$    &500 &1000&500&1000 &250&500&250&500\\
$\left(P_{e^-},\, P_{e^+}\right)$    &(0,\,0) &(0,\,0)&(0,\,0)&(0,\,0) &(.8,\,-.3)&(.8,\,-.3)&(.8,\,.3)&(.8,\,.3)\\
    \hline
 $e^+e^-\to \mu^+\mu^-+\not\!\!E_T $ &0.67 &1.74 &0.55 &1.05 &0.7 &2.02 &0.6 &1.18\\
&&&&&&&&\\ \hline
$e^+e^-\to e^+e^-+\not\!\!E_T $&0.62 &1.64 &0.52 &1 &0.63 &1.67 &0.55 &1.15\\
&&&&&&&&\\ \hline
$e^+e^-\to j\,j+\not\!\!E_T $ &0.98 &2.62 &0.81 &1.53 &1.1 &2.87 &0.89 &1.88\\ 
&&&&&&&&\\ \hline
$e^+e^-\to Z +\not\!\!E_T$ &0.99 &2.64 &0.82 &1.55 &1.1 &2.88 &0.89 &1.89\\
&&&&&&&&\\ \hline\hline
\end{tabular}
\end{center}
\caption{\small \em{ Estimation of  3$\sigma$ sensitivity  reach of the maximum value of the cut-off $\Lambda_{\rm max.}$ that can be probed in ILC  at $\sqrt{s}$ = 500 GeV and 1 TeV   with an integrated luminosity $\mathcal{L} $ = 500 fb$^{-1}$  and $\mathcal{L} $ = 1 ab$^{-1}$  respectively, through all visible channels of $e^+e^-\to Z+ \chi\bar\chi$.}}
\label{table:figureofmerit}
\end{table*}

where $N_{ij}^{NP}$ are number of differential events given by New Physics and $N_{ij}^{SM+NP}$ are total number of differential events for $(ij)^{th}$ grid in double distribution. 
$\delta_{sys}$ is the total systematic error in the measurement. Although the systematic uncertainty inclusive of the luminosity uncertainty  is considered to be of the order of  0.3\% in the literature \cite{Behnke:2013xla}, We have considered a conservative total systematic error to be of order of 1\%. Further, we find that the sensitivity of the cut-off $\Lambda$ can be increased with the increase in the luminosity as the cut-off scales $\sim$ $\mathcal{L}^{1/8}$ for a given $\chi^2$. Thus, for $\sqrt{s}$ = 1 TeV and Luminosity $\mathcal{L}$ = 1 ab$^{-1}$ the sensitivity of the signal will improve by $\approx$ 10 \%. 

\par From the  $\chi^2$ analysis, we obtained 
the $3\,\sigma$ contours in the $m_\chi-\Lambda$ plane corresponding to all the three processes in Figures ~\ref{fig:contour_500_GeV} and ~\ref{fig:contour_1_TeV}. We present the  estimation of  the 3$\sigma$ sensitivity  reach of the maximum value of the cut-off $\Lambda_{\rm max.}$ that can be probed in ILC  at $\sqrt{s}$ =  500 GeV  and 1 TeV  with an integrated luminosity  $\mathcal{L} $ = 500 fb$^{-1}$ and 1 ab$^{-1}$  respectively in Table \ref{table:figureofmerit}. 
\par We compare our results for the unpolarized beams at $\sqrt{s}$ = 500 GeV and ${\cal L}$ = 500 fb$^{-1}$ which are displayed in Figure ~\ref{fig:contour_500_GeV}  with Figures 4d and 4e  of reference  \cite{Yu:2014ula} corresponding to pseudo-scalar and axial-vector couplings respectively. We find that our analysis can provide a better kinematic reach ( $m_{\chi} = 20 $ GeV ) for the pseudo-scalar coupling. We can probe $\Lambda_{\rm max.}$  upto 0.98 TeV, which is further improved to 1.07 TeV with the enhanced integrated luminosity of $\mathcal{L} $ = 1 ab$^{-1}$. However, for the case of axial-vector couplings we can only probe $\Lambda_{\rm max.}$  upto 0.88 TeV as compared to 0.92 TeV given in Figure 4e of \cite{Yu:2014ula}.
\par On the same note, we would like to compare our results for  $\sqrt{s}$ = 1 TeV and ${\cal L}$ = 1 ab$^{-1}$  which are displayed in Figure ~\ref{fig:contour_1_TeV}
with Figures 4d and 4e of reference \cite{Yu:2014ula} corresponding to the pseudo-scalar and axial vector respectively. We find that the sensitivity for the pseudo-scalar can be improved   $\Lambda_{\rm max.}$ = 2.62 TeV in comparison to 2.3 TeV given in Figure 4d of \cite{Yu:2014ula}. Similarly, for axial-vector couplings, we agree with $\Lambda_{\rm max.}$ given in Figure 4e of \cite{Yu:2014ula}.
\subsection{Effect of Polarized Beam}
\label{polarization}
We further study the sensitivity of $\Lambda$ dependence on $m_\chi$ with the polarized initial beams.
The rate of pair production of the fermionic DM through Scalar ($SS$) and PseudoScalar ($PP$) interactions can be enhanced by 
increasing the flux  of  right (left)  handed electrons and right (left) handed positrons. Similarly $VV$ and $AA$ interaction 
of the fermionic DM can be enhanced by choosing the right(left) handed electron beam and left(right) handed positron beam. 
The background contribution from $t$-channel $W$ exchange diagram can be suppressed significantly by choosing right handed polarized 
electron beam which then leaves us with the following choices of polarization combination {\it w.r.t.} helicity conserving($VV$,$AA$) and helicity 
flipping ($SS$,$PP$) interactions respectively
\begin{enumerate}
 \item $+\,80\%$ $e-$ and $-\,30\%$ $e+$
 \item $+\,80\%$ $e-$ and $+\,30\%$ $e+$
\end{enumerate}
\par We exhibit the 99\% C.L. contours corresponding to the polarized initial $e^-$ and $e^+$ beams in Figures \ref{fig:contour_500_GeV} and \ref{fig:contour_1_TeV} in the  $m_\chi$ - $\Lambda$ plane for all possible visible signatures associated with DM pair production at $\sqrt{s}$ = 500 GeV  and 1 TeV with an integrated luminosity of 250 fb$^{-1}$ and 500 fb$^{-1}$ respectively.  
We find that, for scalar  and pseudoscalar (vector and axial vector)  DM couplings, we can probe $\Lambda_{\rm max.}$ values upto 1.1 TeV (0.89 TeV) with polarized initial beams at an integrated Luminosity of 250 fb$^{-1}$.  Similarly, we observe that the analysis with $\sqrt{s}$ = 1 TeV and  an integrated luminosity of 1 ab$^{-1}$ the sensitivity of $\Lambda_{\rm max.}$ can be improved to 3.13 TeV (2.05 TeV) for scalar and pseudoscalar (vector and axial-vector) DM couplings.
\par We perform the combined $\chi^2$ analyses from all the three processes involving the DM pair production for both unpolarized and polarized initial beams. We compute the 99\% confidence limit contours and  plot them in Figures \ref{fig:contour_Comb_500_GeV} and \ref{fig:contour_Comb_1_TeV} for $\sqrt{s}$ = 500 GeV and 1 TeV respectively. 

\subsection{Comparison with the existing analysis}

We present a  comparison of our results  with those that exist in the recent literature.

\subsubsection{Sensitivity from mono-photon channel}
\par First, we make an overall comparison with the other complementary dominant DM production mono-photon channel studied in the reference~\cite{Chae:2012bq}. We have calculated and verified the cross sections and significance of various interactions in the mono-photon channel with the cuts mentioned in~\cite{Chae:2012bq}.
Our analysis shows that the sensitivity of $\Lambda$ is  higher  than that of the mono-photon channel, when mediated by the scalar coupling of the DM, specially in the lower $m_\chi$ region and behave as competitive DM production channel upto  DM mass  $\approx$ 300 GeV. It is important to mention that the enhancement in the sensitivity are obtained  inspite of our  conservative input for the systematic uncertainty $\approx$ 1\% in contrast to that of $\approx$ 0.3\%  in the reference \cite{Chae:2012bq}. 
\begin{table}[h!]\footnotesize
\begin{center}
\begin{tabular}{c||c|c||c|c}
\hline\hline
&\multicolumn{2}{c||}{\bf \underline{Unpolarized}}&\multicolumn{2}{c}{\bf \underline{Polarized}}\\
Interactions    &{V-V} &{A-A} &{V-V} &{A-A} \\
$\left(P_{e^-},\, P_{e^+}\right)$    &(0,\,0) &(0,\,0) &(.8,\,-.3)&(.8,\,-.3)\\
\hline
Reference \cite{Neng:2014mga} &0.96 &0.95 &1.77 &1.76 \\
&&&&\\
Our Analysis &0.99 &0.98  &1.67 &1.66 \\   
&&&&\\ \hline       
\end{tabular}
\end{center}
\caption{\small \em{Efficiency $S/\sqrt{B}$  of the process $e^+e^-\to j\,j+\not\!\!E_T $ for $m_\chi$ =10 GeV and $\Lambda $ = 1 TeV at $\sqrt{s}$ = 500 GeV and  an integrated luminosity ${\cal L}$ = 100 fb$^{-1}$, with unpolarized and polarized initial beams. }}
\label{table:cscom1}
\end{table}

\subsubsection{Results from mono-$Z$ channels}
\par Next, we compare our results with the existing analysis on the DM production channels in association with $Z$ boson which decays to pair of light quarks, electrons and muons. 
\par DM interactions mediated through the vector and axial vector couplings   are compared with the results of the reference \cite{Neng:2014mga} in Table \ref{table:cscom1}. We  marginally improve  the upper limit   $\Lambda_{\rm max}$ ( at $m_\chi$ = 10 GeV) corresponding to the unpolarized  initial beams.   However, it should be noted that the authors of  reference  ~\cite{Neng:2014mga} have assumed $\alpha_{\rm em}^{-1}$ to be  137 at $\sqrt{s}$ = 500 GeV or 1 TeV instead of  127.9,  thus leading to the gross under-estimation of the background  events. Rectifying the error will further bring down the efficiency of the sensitivity of their analysis.

\begin{table}[h!]\footnotesize
\begin{center}
\begin{tabular}{c||c|c||c|c}
\hline\hline
&\multicolumn{2}{c||}{\bf \underline{Unpolarized}}&\multicolumn{2}{c}{\bf \underline{Polarized}}\\

Interactions     &{P-P} &{A-A} &{P-P} &{A-A}\\
     $m_\chi$ in GeV  & 120 & 150 & 120& 150 \\
     $\Lambda$ in GeV & 400 &280 &400 &280\\
$\left(P_{e^-},\, P_{e^+}\right)$    &(0,\,0) &(0,\,0) &(.8,\,-.3)&(.8,\,-.3)\\
\hline
&&&&\\
Reference \cite{Yu:2014ula} &18.7 & 12.3 &34.4 &23.0\\ 
 &&&&\\
Our Analysis &21.13 &18.45 &37.14 &28.74\\     
&&&&\\ \hline       
\end{tabular}
\end{center}
\caption{\small \em{Efficiency $S/\sqrt{S+B}$  of the process $e^+e^-\to j\,j+\not\!\!E_T$ at $\sqrt{s}$ = 500 GeV and an integrated luminosity ${\cal L}$ = 100 fb$^{-1}$ corresponding to the benchmark points mentioned in the  reference \cite{Yu:2014ula} for both unpolarized and polarized initial beams.}}
\label{table:cscom3}
\end{table}
 
\par We also compare the significance of the mono-$Z$ channel processes in the  hadronic decay mode with those given in reference ~\cite{Yu:2014ula} at $\sqrt{s}$ = 500 GeV. We   summarize this in Table~\ref{table:cscom3} for some representative values of   $m_\chi$ and  $\Lambda$.  We find that the our cuts strategy and analysis shows  enhancement in  the signal significance {\it w.r.t.} the  pseudo-scalar and axial-vector interactions of DM with charged leptons for both unpoalrized and  polarized initial beams.



\section{Summary and Outlook}
\label{sec:conclusion}
The fact that a massive ${\cal O}(100 \gev)$ stable particle $\chi$ with
near--weak scale interactions can constitute an attractive candidate
for the Dark Matter in the universe (while satisfying all of competing
constraints from the relic density, large-scale structure formation as
well as myriad other cosmological and astrophysical observations) is
well-known.  While the correct relic abundance can be reproduced by
ensuring the right annihilation cross-sections into any of the SM
particles, direct detection of the DM candidate largely hinges on its
having unsuppressed interactions with the first-generation quarks.

It, then, is of interest to consider the possibility that $\chi$
 has very suppressed (if any) interactions with quarks and
gluons. Not only would the direct-detection experiments have very
little sensitivity to such particles, their pair-production
cross-section at the LHC would be suppressed too. In such circumstances
though, the DM must have significant couplings to at least some of the 
leptons and/or the electroweak gauge bosons. 

\par In this article, we have 
investigated the first of the two possibilities. We obtained the upper bound on the cut-off $\Lambda$ for the given range of electrophilic and/ or leptophilic DM mass $m_{\chi}$  from the relic density constraints which is displayed in the Figure~\ref{fig:relicdensity}. Using these upper bounds on $\Lambda$ for a given $m_\chi$, we estimated the thermally averaged annihilation indirect detection cross-section  and leptophilic DM direct detection scattering cross-section  and are displayed  in Figures ~\ref{fig:indrdetectn} and ~\ref{directdetection} respectively. We find that the present experimental limits in the respective searches not only favours the allowed parameter space from the relic density, but  also constraints the DM  model by providing the lower bound on the  $\Lambda$ for a given DM mass $m_\chi$.

\par If the couplings to the charged leptons be unsuppressed, the linear
collider could play an interesting role in establishing the existence
of such a $\chi$. Using the allowed coupling to the electron, this has,
traditionally, been undertaken using the mono-photon (accompanied by
missing energy-momentum) channel. We have investigated, here, the
complementary channel, namely $e^+ e^- \to \chi \bar \chi + f \bar f$,
where $f$ is any of the light charged fermions (jets, electrons and muons) and exhibited the significance of these processes with basic kinematic cuts in Table \ref{table:cs}.

\par We analysed the sensitivity of DM scalar and vector couplings through the one and two dimensional normalised differential kinematic distributions corresponding to the most dominant signature  $e^+e^-\to j \,j+ \not\!\!E_T$ in Figures \ref{fig:distrSpgs}, \ref{fig:distrVpgs} and \ref{fig:2d} respectively.  99 \% C.L. contours for various sets of run parameters based on the $\chi^2$ analysis of these differential distributions are shown in  Figures ~\ref{fig:contour_500_GeV}, ~\ref{fig:contour_1_TeV}. Combining the $\chi^2$ analysis, from all the three   processes corresponding to $\sqrt{s}$ = 500 GeV and 1 TeV  further improves the sensitivity of the cut-off for given DM mass and are shown in Figures~\ref{fig:contour_Comb_500_GeV} and ~\ref{fig:contour_Comb_1_TeV} respectively. We find that   specific  choice of the initial beam polarisation enhances the sensitivity of the cut-off    for the scalar and vector couplings of DM.    Our analysis shows that the proposed ILC can probe the  cosmologically 
allowed $m_\chi - \Lambda$ region for the leptophilic and/ or electrophilic DM with    higher sensitivity and can provide a direction towards the understanding of the nature of such DM. 

\par We hope this study will be useful in studying the physics potential of the ILC in
context to dark matter searches.

\vskip 5mm
\begin{acknowledgements}
 We would like to thank Debajyoti Cho-udhury for useful discussions 
 and comments on the manuscript. SD acknowledges the partial financial
 support from the CSIR grant No. 03(1340)/15/EMR-II
 BR $\&$ DS acknowledge the UGC for partial financial support. 
 SD and BR would like to thank IUCAA, Pune for hospitality where part of
 this work was completed.
 \end{acknowledgements}


\begin{thebibliography}{99}
\bibitem{Milgrom:1983ca}
M.~{Milgrom}, ``{A modification of the Newtonian dynamics as a possible
  alternative to the hidden mass hypothesis},'' {\em Ap. J.} {\bf 270} (1983)
  365.

\bibitem{galaxy:RubinFord}
V.~C. {Rubin} and W.~K. {Ford}, Jr., ``{Rotation of the Andromeda Nebula from a
  Spectroscopic Survey of Emission Regions},'' {\em Astroph. J.} {\bf 159}
  (1970) 379.
\bibitem{grav_lensing}
L.~A.~Moustakas and R.~B.~Metcalf,
``Detecting dark matter substructure spectroscopically in strong gravitational
lenses,''
Mon.\ Not.\ Roy.\ Astron.\ Soc.\  {\bf 339}, 607 (2003)
[arXiv:astro-ph/0206176].

\bibitem{weaklensing}
  E. van Uitert {\it{et al}} 
  ``Constraints on the shapes of galaxy dark matter haloes from weak gravitational lensing,''
  Astron.\ Astroph.\ {\bf 545} A71(2012)
  [arXiv:1206.4304].

\bibitem{Clowe:2006eq} 
  D.~Clowe, M.~Bradac, A.~H.~Gonzalez, M.~Markevitch, S.~W.~Randall, C.~Jones and D.~Zaritsky,
  ``{\em A direct empirical proof of the existence of dark matter}'',
  Astrophys.\ J.\  {\bf 648}, L109 (2006)
  [astro-ph/0608407].

\bibitem{Hinshaw:2012aka} 
  G.~Hinshaw {\it et al.} [WMAP Collaboration],
  ``{\em Nine-Year Wilkinson Microwave Anisotropy Probe (WMAP) Observations: Cosmological Parameter Results}'',
  Astrophys.\ J.\ Suppl.\  {\bf 208}, 19 (2013)
  [arXiv:1212.5226 [astro-ph.CO]].

\bibitem{Abadi:2002tt}
M.~G. Abadi, J.~F. Navarro, M.~Steinmetz, and V.~R. Eke, ``{Simulations of
  galaxy formation in a lambda CDM universe. 2. The fine structure of simulated
  galactic disks},'' {\em Astrophys. J.} {\bf 597} (2003) 21,
  \href{http://xxx.lanl.gov/abs/astro-ph/0212282}{{\tt astro-ph/0212282}}.

\bibitem{Ade:2015xua} 
  P.~A.~R.~Ade {\it et al.} [Planck Collaboration],
  ``{\em Planck 2015 results. XIII. Cosmological parameters}'',
  arXiv:1502.01589 [astro-ph.CO].

\bibitem{Thomas:2016iav} 
  D.~B.~Thomas, M.~Kopp and C.~Skordis,
  ``{\em Constraining dark matter properties with Cosmic Microwave Background observations}'', 
  arXiv:1601.05097 [astro-ph.CO].

\bibitem{Overduin:2004sz} 
  J.~M.~Overduin and P.~S.~Wesson,
  Phys.\ Rept.\  {\bf 402}, 267 (2004)
  doi:10.1016/j.physrep.2004.07.006
  [astro-ph/0407207].

\bibitem{Hui:2016ltb} 
For example, it has been argued recently that the DM might be 
largely composed of ultralight axions. See, e.g.,\\  L.~Hui, J.~P.~Ostriker, S.~Tremaine and E.~Witten,
  arXiv:1610.08297 [astro-ph.CO].

\bibitem{Moultaka:2006su} 
  G.~Moultaka,
  Acta Phys.\ Polon.\ B {\bf 38}, 645 (2007)
  [hep-ph/0612331].

\bibitem{Dutta:2015ega} 
  S.~Dutta, A.~Goyal and S.~Kumar,
  JCAP {\bf 1602}, no. 02, 016 (2016)
  doi:10.1088/1475-7516/2016/02/016
  [arXiv:1509.02105 [hep-ph]].

\bibitem{Davoudiasl:2017jke} 
  H.~Davoudiasl and C.~W.~Murphy,
  Phys.\ Rev.\ Lett.\  {\bf 118}, no. 14, 141801 (2017)
  doi:10.1103/PhysRevLett.118.141801
  [arXiv:1701.01136 [hep-ph]].

\bibitem{Feng:2010gw} 
  J.~L.~Feng,
  Ann.\ Rev.\ Astron.\ Astrophys.\  {\bf 48}, 495 (2010)
  doi:10.1146/annurev-astro-082708-101659
  [arXiv:1003.0904 [astro-ph.CO]].

\bibitem{Bernabei:2013xsa} 
  R.~Bernabei {\it et al.},
  ``{\em Final model independent result of DAMA/LIBRA-phase1}'', 
  Eur.\ Phys.\ J.\ C {\bf 73}, 2648 (2013)
  [arXiv:1308.5109 [astro-ph.GA]].

\bibitem{Aalseth:2012if} 
  C.~E.~Aalseth {\it et al.} [CoGeNT Collaboration],
  Phys.\ Rev.\ D {\bf 88}, 012002 (2013)
  doi:10.1103/PhysRevD.88.012002
  [arXiv:1208.5737 [astro-ph.CO]].

\bibitem{Angloher:2016rji} 
  G.~Angloher {\it et al.} [CRESST Collaboration],
  arXiv:1612.07662 [hep-ex].


\bibitem{Aprile:2016wwo} 
  E.~Aprile {\it et al.} [XENON Collaboration],
  Phys.\ Rev.\ D {\bf 94}, no. 9, 092001 (2016)
  doi:10.1103/PhysRevD.94.092001
  [arXiv:1605.06262 [astro-ph.CO]].

\bibitem{Yang:2016odq} 
  Y.~Yang [PandaX-II Collaboration],
  arXiv:1612.01223 [hep-ex].

\bibitem{Tan:2016zwf} 
  A.~Tan {\it et al.} [PandaX-II Collaboration],
  Phys.\ Rev.\ Lett.\  {\bf 117}, no. 12, 121303 (2016)
  doi:10.1103/PhysRevLett.117.121303
  [arXiv:1607.07400 [hep-ex]].

\bibitem{Akerib:2016vxi} 
  D.~S.~Akerib {\it et al.},
  arXiv:1608.07648 [astro-ph.CO].

\bibitem{Fermi-LAT:2016uux} 
  A.~Albert {\it et al.} [Fermi-LAT and DES Collaborations],
  Astrophys.\ J.\  {\bf 834}, no. 2, 110 (2017)
  doi:10.3847/1538-4357/834/2/110
  [arXiv:1611.03184 [astro-ph.HE]].

\bibitem{Adriani:2008zr} 
  O.~Adriani {\it et al.} [PAMELA Collaboration],
  Nature {\bf 458}, 607 (2009)
  doi:10.1038/nature07942
  [arXiv:0810.4995 [astro-ph]].

\bibitem{Aguilar:2013qda} 
  M.~Aguilar {\it et al.} [AMS Collaboration],
  Phys.\ Rev.\ Lett.\  {\bf 110}, 141102 (2013).
  doi:10.1103/PhysRevLett.110.141102

 \bibitem{Bhattacherjee:2012ch} 
  B.~Bhattacherjee, D.~Choudhury, K.~Harigaya, S.~Matsumoto and M.~M.~Nojiri,
  ``{\em Model Independent Analysis of Interactions between Dark Matter and Various Quarks}'',
  JHEP {\bf 1304}, 031 (2013)
  [arXiv:1212.5013 [hep-ph]].
  

\bibitem{Goodman:2010ku}
J.~Goodman, M.~Ibe, A.~Rajaraman, W.~Shepherd, T.~M. Tait, et~al., ``{\em
  Constraints on Dark Matter from Colliders}'',  {\em Phys.Rev.} {\bf D82} 116010 (2010) [arXiv:1008.1783].
  
  
\bibitem{Carpenter:2012rg}
L.~M. Carpenter, A.~Nelson, C.~Shimmin, T.~M. Tait, and D.~Whiteson, {\em Collider searches for dark matter in events with a Z boson and missing
  energy},  {\em Phys.Rev.} {\bf D87} no.~7 074005 (2013) ,
  [arXiv:1212.3352 [hep-ph]].-----

\bibitem{Petrov:2013nia}
A.~A. Petrov and W.~Shepherd, {\em Searching for dark matter at LHC with
  Mono-Higgs production},  {\em Phys.Lett.} {\bf B730} 178--183 (2014) ,
  [arXiv:1311.1511 [hep-ph]].

\bibitem{Carpenter:2013xra}
L.~Carpenter, A.~DiFranzo, M.~Mulhearn, C.~Shimmin, S.~Tulin, et~al., {\em Mono-Higgs-boson: A new collider probe of dark matter},  {\em Phys.Rev.}
  {\bf D89} no.~7 075017, (2014), [arXiv:1312.2592 [hep-ph]].

\bibitem{Berlin:2014cfa}
A.~Berlin, T.~Lin, and L.-T. Wang, {\em Mono-Higgs Detection of Dark Matter at
  the LHC},  {\em JHEP} {\bf 1406} 078 (2014),
  [arXiv:1402.7074 [hep-ph]].

\bibitem{Lin:2013sca}
T.~Lin, E.~W. Kolb, and L.-T. Wang, {\em Probing dark matter couplings to top
  and bottom quarks at the LHC},  {\em Phys.Rev.} {\bf D88}, no.~6
  063510, (2013), [arXiv:1303.6638 [hep-ph]].


\bibitem{Bai:2010hh}
Y.~Bai, P.~J. Fox, and R.~Harnik, ``{\em The Tevatron at the Frontier of Dark
  Matter Direct Detection}'',  {\em JHEP} {\bf 1012} 048 (2010),
   [arXiv:1005.3797].

\bibitem{Birkedal:2004xn}
A.~Birkedal, K.~Matchev, and M.~Perelstein, ``{\em Dark matter at colliders: A
  Model independent approach}'',  {\em Phys.Rev.} {\bf D70} 077701 (2004),
  { hep-ph/0403004}].

\bibitem{Gershtein:2008bf}
Y.~Gershtein, F.~Petriello, S.~Quackenbush, and K.~M. Zurek, {\em Discovering
  hidden sectors with mono-photon $Z^\prime$ searches},  {\em Phys.Rev.} {\bf
  D78} 095002 (2008) , [arXiv:0809.2849 [hep-ph]].

\bibitem{Feng:2005gj} 
  J.~L.~Feng, S.~Su and F.~Takayama,
  Phys.\ Rev.\ Lett.\  {\bf 96}, 151802 (2006)
  doi:10.1103/PhysRevLett.96.151802
  [hep-ph/0503117].

\bibitem{Fox:2011fx}
P.~J. Fox, R.~Harnik, J.~Kopp, and Y.~Tsai, {\em LEP Shines Light on Dark
  Matter},  {\em Phys.Rev.} {\bf D84} 014028 (2011),
  [arXiv:1103.0240 [hep-ph]].
\bibitem{Bai:2012xg}
Y.~Bai and T.~M. Tait, {\em Searches with Mono-Leptons},  {\em Phys.Lett.}
  {\bf B723} 384--387 (2013),   
[arXiv:1208.4361 [hep-ph]].

\bibitem{Crivellin:2015wva}
A.~Crivellin, U.~Haisch, and A.~Hibbs, {\em LHC constraints on gauge boson
  couplings to dark matter},[arXiv:1501.00907].

\bibitem{Petriello:2008pu}
F.~J. Petriello, S.~Quackenbush, and K.~M. Zurek, {\em The Invisible
  $Z^\prime$ at the CERN LHC},  {\em Phys.Rev.} {\bf D77} 115020 (2008),
  [arXiv:0803.4005 [hep-ph]].

\bibitem{Chae:2012bq}
  Y.~J.~Chae and M.~Perelstein,
  ``{\em Dark Matter Search at a Linear Collider: Effective Operator Approach}'', 
  JHEP {\bf 1305} (2013) 138
  [arXiv:1211.4008 [hep-ph]].

\bibitem{Dreiner:2012xm} 
  H.~Dreiner, M.~Huck, M.~Krämer, D.~Schmeier and J.~Tattersall,
  Phys.\ Rev.\ D {\bf 87}, no. 7, 075015 (2013)
  [arXiv:1211.2254 [hep-ph]].

\bibitem{Yu:2013aca} 
  Z.~H.~Yu, Q.~S.~Yan and P.~F.~Yin,
  Phys.\ Rev.\ D {\bf 88}, no. 7, 075015 (2013)
  doi:10.1103/PhysRevD.88.075015
  [arXiv:1307.5740 [hep-ph]].

\bibitem{Bartels:2007cv} 
  C.~Bartels and J.~List,
  eConf C {\bf 0705302}, COS02 (2007)
  [arXiv:0709.2629 [hep-ex]].

\bibitem{Griest:1990kh} 
  K.~Griest and D.~Seckel,
  ``{\em Three exceptions in the calculation of relic abundances}'',
  Phys.\ Rev.\ D {\bf 43}, 3191 (1991).
  


\bibitem{Zheng:2010js} 
  J.~M.~Zheng, Z.~H.~Yu, J.~W.~Shao, X.~J.~Bi, Z.~Li and H.~H.~Zhang,
  ``{\em Constraining the interaction strength between dark matter and visible matter: I. fermionic dark matter}'',
  Nucl.\ Phys.\ B {\bf 854}, 350 (2012)
  [arXiv:1012.2022 [hep-ph]].
  
 \bibitem{Fixsen:2009ug} 
  D.~J.~Fixsen,
  Astrophys.\ J.\  {\bf 707}, 916 (2009)
  [arXiv:0911.1955 [astro-ph.CO]].
  \bibitem{Bennett:2012zja} 
  C.~L.~Bennett {\it et al.}  [WMAP Collaboration],
  Astrophys.\ J.\ Suppl.\  {\bf 208}, 20 (2013)
  [arXiv:1212.5225 [astro-ph.CO]].
  
  
\bibitem{Christensen:2008py} 
  N.~D.~Christensen and C.~Duhr,
  Comput.\ Phys.\ Commun.\  {\bf 180}, 1614 (2009)
  [arXiv:0806.4194 [hep-ph]].
  
\bibitem{Belanger:2001fz} 
  G.~Belanger, F.~Boudjema, A.~Pukhov and A.~Semenov,
  Comput.\ Phys.\ Commun.\  {\bf 149}, 103 (2002)
  doi:10.1016/S0010-4655(02)00596-9
  [hep-ph/0112278].

\bibitem{Backovic:2013dpa} 
  M.~Backovic, K.~Kong and M.~McCaskey,
  Physics of the Dark Universe {\bf 5-6}, 18 (2014)
  doi:10.1016/j.dark.2014.04.001
  [arXiv:1308.4955 [hep-ph]].

\bibitem{Backovic:2015tpt} 
  M.~Backovic, A.~Martini, K.~Kong, O.~Mattelaer and G.~Mohlabeng,
  AIP Conf.\ Proc.\  {\bf 1743}, 060001 (2016)
  doi:10.1063/1.4953318
  [arXiv:1509.03683 [hep-ph]].
  \bibitem{Kopp:2009et} 
  J.~Kopp, V.~Niro, T.~Schwetz and J.~Zupan,
  ``{\em DAMA/LIBRA and leptonically interacting Dark Matter}'', 
  Phys.\ Rev.\ D {\bf 80}, 083502 (2009)
  [arXiv:0907.3159 [hep-ph]].

\bibitem{Aprile:2015ade} 
  E.~Aprile {\it et al.} [XENON100 Collaboration],
  ``{\em Exclusion of Leptophilic Dark Matter Models using XENON100 Electronic Recoil Data}'',
  Science {\bf 349}, no. 6250, 851 (2015)
  doi:10.1126/science.aab2069
  [arXiv:1507.07747 [astro-ph.CO]].
  
\bibitem{Neng:2014mga} 
  W.~Neng, S.~Mao, L.~Gang, M.~Wen-Gan, Z.~Ren-You and G.~Jian-You,
  Eur.\ Phys.\ J.\ C {\bf 74}, no. 12, 3219 (2014),
  [arXiv:1403.7921 [hep-ph]].
  
\bibitem{Yu:2014ula}
Z.-H. Yu, X.-J. Bi, Q.-S. Yan, and P.-F. Yin, {\em Dark matter searches in the
  mono-$Z$ channel at high energy $e^+e^-$ colliders},  {\em Phys.Rev.} {\bf
  D90} no.~5 055010, (2014), [arXiv:1404.6990 [hep-ph]].

\bibitem{Alwall:2014hca} 
  J.~Alwall {\it et al.},
  JHEP {\bf 1407}, 079 (2014)
  doi:10.1007/JHEP07(2014)079
  [arXiv:1405.0301 [hep-ph]].
  
\bibitem{Sjostrand:2006za}
  T.~Sjostrand, S.~Mrenna and P.~Z.~Skands,
  JHEP {\bf 0605}, 026 (2006)
  [hep-ph/0603175].

\bibitem{pgs}
  PGS-4, J. Conway et al.,
  \url{http://www.physics.ucdavis.edu/~conway/research/software/pgs/pgs4-general.htm}.


\bibitem{Behnke:2013lya} 
  T.~Behnke {\it et al.},
  arXiv:1306.6329 [physics.ins-det].


\bibitem{Behnke:2013xla} 
  T.~Behnke {\it et al.},
  ``The International Linear Collider Technical Design Report - Volume 1: Executive Summary,''
  arXiv:1306.6327 [physics.acc-ph].
 
\bibitem{Conte:2012fm} 
  E.~Conte, B.~Fuks and G.~Serret,
  Comput.\ Phys.\ Commun.\  {\bf 184}, 222 (2013)
  doi:10.1016/j.cpc.2012.09.009
  [arXiv:1206.1599 [hep-ph]].

\end{thebibliography}
\end{document}